%Paper: hep-th/9403148
%From: juriev@physique.ens.fr (JURIEV Denis)
%Date: Thu, 24 Mar 94 13:29:43 +0100
%Date (revised): Mon, 28 Mar 94 10:14:28 +0200

% This is AMSTEX.
\input amstex
\documentstyle{amsppt}
\NoRunningHeads
\NoBlackBoxes
 \define\FA{\frak A} \define\CA{\Cal A}
  \define\CB{\Cal B}
  \define\CC{\Cal C}
  
\define\Fe{\frak e}  
  \define\CF{\Cal F}
\define\Fg{\frak g} \define\FG{\frak G} 
  
\define\Fi{\frak i}

  \define\CL{\Cal L}
  
  \define\CN{\Cal N}
  \define\CO{\Cal O}
  
  \define\CQ{\Cal Q}
  \define\CR{\Cal R}
  \define\CS{\Cal S}
\define\Ft{\frak t}  
  \define\CU{\Cal U}
  
\define\Fw{\frak w}

\define\Fz{\frak z}  
%-----

\define\AC{\Bbb C}

   \define\BI{\bold I}

\define\AP{\Bbb P}   \define\BP{\bold P} \define\Bp{\bold p}
\define\AQ{\Bbb Q}   
\define\AR{\Bbb R}   \define\BR{\bold R}
\define\AS{\Bbb S}

\define\AZ{\Bbb Z}   
%-----

\define\TY{\text{\tt Y}}
%-----
 \define\vtheta\vartheta
 
\define\ga{\alpha}
\define\gb{\beta}
 \redefine\gg{\gamma}  
\define\gd{\delta} \define\gD{\Delta}
 
\define\gz{\zeta}

\define\gl{\lambda}

%%%%%%%%%%%%%%%%%%%%%%%%%%%%%%%%%%%%%%%%%%%%%%%%%%%%%%%%%%%%%%%%%%%%%%%%%%
\define\Hom{\operatorname{Hom}}
\define\Mor{\operatorname{Mor}}

\define\Der{\operatorname{Der}}

\define\id{\operatorname{id}}

\define\ad{\operatorname{ad}}

\define\Mat{\operatorname{Mat}}
\define\Map{\operatorname{Map}}
%-------------------------------------------------------------------------
\define\pd#1#2{\partial_{#1}^{#2}}
\define\p#1{\partial_{#1}}
\define\fpd#1#2{(\frac{\partial}{\partial {#1}})^{#2}}
\define\fp#1{\frac{\partial}{\partial {#1}}}
%-------------------------------------------------------------------------
\define\Gr#1#2#3{#1(#2,#3)}
\define\GrC#1#2{\Gr#1#2{\AC}}
\define\GrR#1#2{\Gr#1#2{\AR}}
\define\GrQ#1#2{\Gr#1#2{\AQ}}
\define\GrZ#1#2{\Gr#1#2{\AZ}}
\define\vGr#1#2#3{#1_{#2}(#3)}
\define\vGrC#1#2{\vGr#1#2{\AC}}
\define\vGrR#1#2{\vGr#1#2{\AR}}
\define\vGrQ#1#2{\vGr#1#2{\AQ}}
\define\vGrZ#1#2{\vGr#1#2{\AZ}}
%-----

%-----

\define\res{\operatorname{res}}
\define\nin{\overline{\in}}
\define\Cdot{\operatorname{\cdot}}
\define\NS{\operatorname{\CN\CS}}
\define\Vrt{\operatorname{Vert}}
\define\Cvir{\operatorname{\AC vir}}
\define\Vect{\operatorname{Vect}}
\define\CVect{\operatorname{\AC Vect}}
\define\LDiff{\operatorname{\CL Diff}}
\define\Diff{\operatorname{Diff}}
\define\circle{\AS^1}
\define\vir{\operatorname{vir}}
\define\Vir{\operatorname{Vir}}
\define\Rot{\operatorname{Rot}}
\define\Ner{\operatorname{Ner}}
\define\Mantle{\operatorname{Mantle}}
\define\Voile{\operatorname{Voile}}
\define\Train{\operatorname{Train}}
\define\Trinion{\operatorname{Trinion}}
\define\Polynion{\operatorname{Polynion}}
\define\sdp#1{\sssize{\frac{\partial}{\partial #1}}\tsize}
\define\inp{\operatorname{in}}
\define\outp{\operatorname{out}}
\define\sLtwo{\operatorname{sl(2,\AC)}}
\leftheadtext{D.JURIEV}
\document

\

\

\

\

\

\topmatter
\title INFINITE DIMENSIONAL GEOMETRY AND QUANTUM FIELD THEORY OF
STRINGS\linebreak
\eightpoint II. INFINITE DIMENSIONAL NONCOMMUTATIVE GEOMETRY OF
A SELF--INTERACTING STRING FIELD
\endtitle
\author
\centerline{}
\centerline{\eightpoint D.JURIEV\footnote{\rm On leave from Mathematical
Division, Research Institute for System Studies, Russian Academy of Sciences,
Moscow, Russia (e mail: juriev\@systud.msk.su).\newline}}
\centerline{}
{\eightpoint\it Laboratoire de Physique Th\'eorique
de l'\'Ecole Normale Sup\'erieure,\linebreak
rue Lhomond 24, 75231 Paris, France\footnote{\rm Unit\'e Propre du Centre
National
de la Recherche Scientifique associ\'ee \`a l'\'Ecole Normale Sup\'erieure et
\`a l'Universit\'e de Paris--Sud.\newline}}\linebreak
\centerline{\eightpoint\rm E mail: juriev\@physique.ens.fr}
\endauthor
\abstract A geometric interpretation of quantum self--interacting
string field theory is given. Relations between various approaches to the
second quantization of an interacting string are described in terms of the
geometric quantization. An algorythm to construct a quantum nonperturbative
interacting string field theory in the quantum group formalism is proposed.
Problems of a metric background (in)dependence are discussed.
\endabstract
\endtopmatter
\tenpoint
This is the second part of the paper devoted to various structures of an
infinite dimensional geometry, appearing in quantum field theory of (closed)
strings; the objects connected with the second quantization of a free string
were described in the first part [1], whereas an analogous material for a
self--interacting string field will be discussed now.

It is proposed to continue the investigation of an infinite dimensional
geometry related to the quantum field theory of strings in the following two
parts (parts III,IV). The third part is devoted to an infinite dimensional
$W$--geometry of a second quantized free string [2]; the fourth part should
contain materials on infinite dimensional geometry of a self--interacting
$W$--string field.

In the whole paper we follow a general ideology of string theory presented in
[3].  All four parts of the publication maybe considered as a sequel of
previous one [4] devoted to geometric aspects of quantum conformal field
theory: the transition from the 2D quantum conformal field theory to the
self--interacting string field theory maybe considered as one from the
abstract geometry of noncommutative Riemann surfaces (in spirit of [4]) to
aspects of, roughly speaking, their "imbeddings" into target spaces. Such
geometric picture seems to underlie a very powerful machinery for explorations
of geometry of those spaces (see [5] for some topological parallels as well as
[6] on so-called "quantum cohomology" as an essential part of them).  The
naturality of the framework of the infinite dimensional geometry for the
subject was clearly explaned in [7,4]. Here it should be only specially
mentioned that in the second quantized formalism the imbeddings of algebraic
curves into target spaces are described by algebraic structures on the space
of their germs at initial point, which encode all global topological or
geometric information on the intrinsic as well as the extrinsic geometry of a
world-sheet (cf. with similar situation in the theory of univalent functions,
where one predicts global behaviour of a conformal mapping by coefficients of
its Taylor expansion [8]); it allows to account the nonperturbative effects,
which seems to be essential for differential geometry in contrast to topology
(in the first quantized approach such effects maybe described only by
something cumbersome as imbeddings of Riemann surfaces of infinite genus).
Thus, a transition from the first quantized formalism to the second quantized
one means an enlargement of the substantially finite-dimensional
algebro-geometric picture [9] by infinite dimensional geometric one. It
produce a lot of questions, which should be attributed to the functional
analysis, nevertheless, we try to avoid them below if they do not explicate
straightforwardly the underlying infinite dimensional geometry (it does not
mean that they are regarded as less important mathematically, but only as
"second order" ones for our rather complicated geometric picture).

The author is very glad to thank Prof.J.-L.Gervais and
Laboratoire de Physique Th\'eorique de l'\'Ecole Normale Sup\'erieure (Paris)
for a warm atmosphere and a kind hospitality, without which this work would
never be finished.

\head 1. Infinite dimensional noncommutative geometry and\linebreak
2D quantum conformal field theory
\endhead

In this chapter the material of [4] is exposed with a lot of new details. The
main idea is that all principal structures of 2D QCFT maybe related to some
faces of the noncommutative geometry [10,11]. The categoric (Kontsevich--Segal)
approach is described in par.1.1. The approach based on operator algebras is
described in par.1.2. as an infinitesimal counterpart of the first. Par.1.3.
is devoted to a description of the renormalization of pointwise product in
operator algebras and, therefore, an imbedding of both approaches into a
framework of the noncommutative geometry. Relations of the noncommutative
geometrical formulation of 2D QCFT with the Krichever--Novikov formalism (in
its renormalized version) are also considered.

\subhead 1.1. The Lie algebra $\Vect(\circle)$ of vector fields on a circle
$\circle$, the group $\Diff_+(\circle)$ of diffeomorphisms of a circle
$\circle$, the Virasoro algebra $\vir$, the Virasoro--Bott group $\Vir$ and
the Neretin semigroup $\Ner$ (the mantle $\Mantle(\Diff_+(\circle))$ of the
group of diffeomorphisms of a circle); the semigroup
$\Voile(\Diff_+(\circle))$ -- the voile of the group of diffeomorphisms of a
circle, manifolds $\Trinion(\Diff_+(\circle))$ and
$\Polynion(\Diff_+(\circle))$ of trinions and polynions, the Kontsevich--Segal
category $\Train(\Diff_+(\circle))$ -- the train of the group of
diffeomorphisms of a circle, the modular functor \endsubhead Let us briefly
expose some facts and constructions following [1,4]. Let $\Diff(\circle)$ be
the group of analytic diffeomorphisms of a circle $\circle$, it consists of
two connected components -- the normal subgroup $\Diff_+(\circle)$ of
diffeomorphisms preserving an orientation on a circle $\circle$ and the coset
$\Diff_-(\circle)$ of ones changing it; a Lie algebra of the connected group
$\Diff_+(\circle)$ is identified with the Lie algebra $\Vect(\circle)$ of
analytic vector fields $v(t)d/dt$ on a circle $\circle$; the structure
constants of the complexification $\CVect(\circle)$ of the Lie algbera
$\Vect(\circle)$ have the form $c^l_{jk}=(j-k)\delta^l_{j+k}$ in the basis
$e_k:=i\cdot\exp(ikt)d/dt$; in 1968 I.M.Gelfand and D.B.Fuchs discovered a
non--trivial central extension of Lie algebra $\Vect(\circle)$, the
corresponding 2--cocycle maybe written as $c(u,v)=\int u'(t)dv'(t)$;
independently this central extension was discovered in 1969 by M.Virasoro and
was called later {\it the Virasoro algebra\/} $\vir$ (the same name belongs to
the complexification $\Cvir$ of this algebra), the Virasoro algebra $\Cvir$ is
generated by the vectors $e_k$ and the central element $c$, the commutation
relations in it have the form
$[e_j,e_k]=(j-k)e_{j+k}+\delta(j+k)\cdot\frac{j^3-j}{12}\cdot c$; one may
correspond an infinite dimensional group $\Vir$ to the Lie algebra $\vir$
which is a central extension of the group $\Diff_+(\circle)$, the
corresponding 2--cocycle was calculated by R.Bott in 1977
$c(f,g)=\int\log(g(f))'d\log f'$, the group $\Vir$ is called {\it the
Virasoro--Bott group}. There are no any infinite dimensional groups
corresponding to the Lie algebras $\CVect(\circle)$ and $\Cvir$ but it is
useful to consider the following construction, which is attributed to
Yu.Neretin and was developped by M.Kontsevich: let us consider accordingly to
Yu.Neretin a local group $\LDiff^{\AC}_+(\circle)$ of all analytic mappings
$g$ from $\circle$ to\linebreak $\AC\backslash\{0\}$ with a Jordan image
$g(\circle)$ homotopical to $\circle$ as an oriented contour in $\Bbb
C\backslash\{0\}$ such that $g'(e^{it})$ is not equal to $0$ anywhere; the
Neretin semigroup $\Ner$ is just the semigroup of all elements $g$ of
$\LDiff^{\AC}_+(\circle)$ such that $|g(\exp(it))|<1$, it is evident that
$\Ner$ is a local semigroup, the globalization was performed by Yu.Neretin (a
brief description of Yu.Neretin construction one may find in [1,7], the
original and more expanded version is contained in [12,13]). As it was shown
by M.L.Kontsevich the elements of {\it the Neretin semigroup\/} $\Ner$ maybe
identified with triples $(K,p,q)$, where $K$ is a Riemann surface with a
boundary biholomorphically equivalent to a ring, $p$ and $q$ are fixed
analytic parametrizations of the two components of the boundary $\partial K$
of the surface $K$, so that $p$ is an input parametrization and $q$ is an
output parametrization; the product of two elements of the Neretin semigroup
is just the gluing of Riemann surfaces. The Neretin semigroup $\Ner$ is called
{\it the mantle\/} of the group $\Diff_+(\circle)$ of diffeomorphisms of a
circle and is denoted also by $\Mantle(\Diff_+(\circle))$; the Neretin
semigroup $\Ner$ admits a central extension $\widehat{\Ner}$, the
corresponding 2--cocycle was calculated by Yu.Neretin in 1989.

Following the way of [14,App.3] it is reasonable to consider a certain
generalization of the mantle $\Mantle(\Diff_+(\circle))$ -- {\it the voile\/}
$\Voile(\Diff_+(\circle))$ of the group of diffeomorphisms of a circle. The
elements of the voile $\Voile(\Diff_+(\circle))$ are triples $(K,p,q)$, where
$K$ is a Riemann surface of arbitrary genus with two component boundary, $p$
and $q$ are fixed analytic parametrizations of these components, $p$ is an
input parametrization and $q$ is an output parametrization; the product of two
elemnts of the semigroup $\Voile(\Diff_+(\circle))$ is the sewing of Riemann
surfaces. The genus of Riemann surfaces defines a system of characters of the
voile $\Voile(\Diff_+(\circle))$ as well as its grading. As it was marked in
[14,App.3] it is very interesting to consider the "fractal" completion of the
voile $\Voile(\Diff_+(\circle))$ including surfaces of infinite genus, which
maybe considered as a "fluctuating" exponent of the Lie algebra
$\CVect(\circle)$ describing an evolution of virtual particles' clothes.

The semigroup constructions (related to the mantle $\Mantle(\Diff_+(\circle))$
and the voile $\Voile(\Diff_+(\circle))$ of the group of diffeomorphisms of a
circle) maybe generalized to the category constructions (related to {\it the
train\/} $\Train(\Diff_+(\circle))$ of the group of diffeomorphisms of a
circle) following to G.Segal [15].  Nevertheless, for the following purposes
it is rather resonable to consider constructions of manifolds
$\Trinion(\Diff_+(\circle))$ and $\Polynion(\Diff_+(\circle))$ of trinions and
polynions and their representations before the general exposition of ones of
the Kontsevich--Segal category $\Train(\Diff_+(\circle))$ and its
representations (modular functors) by an analogy to [14,App.3].

First of all let's briefly remind a geometric way to construct the highest
weight representations over the Virasoro algebra $\Cvir$ [16] and their
extensions to projective representations of the Neretin semigroup $\Ner$
[12,13] based on the infinite dimensional geometry of the flag manifold for
the Virasoro--Bott group [17,16,7]. The flag manifold $M(\Vir)$ for the
Virasoro--Bott group is the homogeneous space $\Diff_+(\circle)/\circle$;
there exist several realizations of this manifold: the realization of
$M(\Vir)$ as an infinite dimensional homogeneous space
$\Diff_+(\circle)/\circle$ is called algebraic (in this realization $M(\Vir)$
maybe identified also with the quotient of the Neretin semigroup $\Ner$ by its
subsemigroup $\Ner^{\circ}$ consisting of elements $g\in\Ner$, which admit an
analytic extension to $D_-=\{z\in\AC: |z|\ge 1\}$); in the probabilistic
realization the group $\Diff_+(\circle)$ acts on the space of all
probabilistic measures $u(t)\,dt$ on $\circle$ with an analytic positive
density $u(t)$ in a natural way; in the orbital realization the space
$M(\Vir)$ is identified with coadjoint orbits of the groups $\Diff_+(\circle)$
and $\Vir$ (such realization provides $M(\Vir)$ by a two--parametric family of
symplectic structures $\omega_{h,c}$); in the analytic realization the space
$M(\Vir)$ is identified with the class $S$ of functions $f(z)$ analytic and
univalent in the closed unit disc $D_+$ normalized by the conditions $f(0)=0$,
$f'(0)=1$, $f'(\exp(it))\ne 0$ by {\it the Kirillov construction\/} (the
Taylor coefficients $c_1, c_2, c_3,\ldots c_k,\ldots$ of a function
$f(z)=z+c_1z^2+c_2z^3+\ldots+c_kz^{k+1}+\ldots$ determine a coordinate system
on $S$; necessary and sufficient conditions for univalency of a function
$f(z)$ maybe found in [8]; the action of $\CVect(\circle)$ on $M(\Vir)$ has
the form $\CL_vf(z)=-if^2(z)\oint\sssize\left[\frac{wf'(w)}{f(w)}\right]^2
\frac{v(w)}{f(w)-f(z)}\frac{dw}w$ ($f\!\in\!S$, $v\!\in\!\CVect(\circle)$).
The symplectic structures $\omega_{h,c}$ coupled with complex structure on
$M(\Vir)$ form the two--parametric family of (pseudo)--K\"ahler metrics
$w_{h,c}$;  a geometric way to construct the Verma modules over the Virasoro
algebra is based on the following facts: (1) To each
$\Diff_+(\circle)$--invariant K\"ahler metric $w_{h,c}$ $w_{h,c}$ on the space
$M(\Vir)$ one should correspond the linear holomorphic bundle $E_{h,c}$ over
$M(\Vir)$ with the following properties:  (a) $E_{h,c}$ is the Hermitean
bundle with metric $\exp(-U_{h,c})\,d\gl\,d\bar{\gl}$, where $\gl$ is a
coordinate in a fiber, $K_{h,c}=\exp(U_{h,c})$ is the Bergman kernfunction,
the exponential of the K\"ahler potential of the metric $w_{h,c}$, (b) algebra
$\Cvir$ holomorphically acts in the prescribed bundle by covariant derivatives
with respect to the hermitean connection with the curvature form being equal
to $2\pi i\omega_{h,c}$; (2) let $\CO(E_{h,c})$ be the space of all polynomial
(in some natural trivialization) germs of sections of the bundle $E_{h,c}$
(the action of $\Cvir$ in its $\AZ_+$--graded module $\CO(E_{h,c})$
($\deg(c_k)=k$) is defined by the formulas $L_p=\sdp{c_p}+\sum_{\sssize{k\ge
1}\tsize}(k+1)c_k\sdp{c_{k+p}}$ ($p>0$), $L_0=\sum_{\sssize{k\ge
1}\tsize}kc_k\sdp{c_k}+h$, $L_{-1}=\sum_{\sssize{k\ge
1}\tsize}((k+2)c_{k+1}\!-\!2c_1c_k)\sdp{c_k}+2hc_1$,
$L_{-2}=\sum_{\sssize{k\ge
1}\tsize}((k+3)c_{k+2}\!-\!(4c_2\!-\!c_1^2)c_k\!-\!b_k(c_1,\ldots
c_{k+2}))\sdp{c_k}+h(4c_2\!-\!c_1^2)+{\sssize{\frac{c}2}}(c_2\!-\!c_1^2)$,
$L_{-n}={\sssize{\frac1{(n-2)!}}}\ad^{n-2}L_{-1}\cdot L_{-2}$ ($n>0$)); let us
fix the basis $e^{a_1,\ldots a_n}=c_1^{a_1}\ldots c_n^{a_n}$ in
$\CO(E^*_{h,c})$ and let $\CO^*(E^*_{h,c})$ be the space of all linear
functionals $p$ on $\CO(E^*_{h,c})$, which obey the property: if $p(x)=0$ then
$\deg(x)\le N_p$ (the space $\CO^*(E^*_{h,c})$ is called the Fock space of the
pair ($M(\Vir)$,$E_{h,c}$) and is denoted by $F(E_{h,c})$); the Verma module
$V_{h,c}$ over $\Cvir$ is realized in the Fock space $F(E_{h,c})$ and if one
fix the basis $e_{a_1,\ldots a_n}=: c_1^{a_1}\ldots c_n^{a_n}$ in $F(E_{h,c})$
such that $<e_{a_1,\ldots a_n},e^{b_1,\ldots b_m}>=a_1!\ldots
a_n!\,\gd^m_n\gd_{a_1}^{b_1}\ldots\gd_{a_n}^{b_n}$ the action of the Virasoro
algebra in such basis will be defined by the formulas
$L_{-p}=c_p+\sum_{\sssize{k\ge 1}\tsize}c_{k+p}\sdp{c_k}$ ($p>0$),
$L_0=\sum_{\sssize{k\ge 1}\tsize} kc_k\sdp{c_k}+h$, $L_1=\sum_{\sssize{k\ge
1}\tsize} c_k((k+2)\sdp{c_{k+1}}\!-\!2\sdp{c_1}\sdp{c_k}) +2h\sdp{c_1}$,
$L_2=\sum_{\sssize{k\ge 1}\tsize}
c_k((k+3)\sdp{c_{k+2}}\!-\!(4\sdp{c_2}\!-\!(\sdp{c_1})^2)\sdp{c_k}\!-\!
b_k(\sdp{c_1},\ldots\sdp{c_{k+2}})) +h(4\sdp{c_2}\!-\!(\sdp{c_1})^2)+
{\sssize{\frac {c}2}}(\sdp{c_2}\!-\!(\sdp{c_1})^2)$,
$L_n={\sssize{\frac{(-1)^n}{(n-2)!}}}\ad^{n-2}L_1\cdot L_2$ ($n>2$).  Such
action of $\Cvir$ maybe exponentiated to a projective representation of
$\Ner$.

Points of a manifold $\Trinion(\Diff_+(\circle))$ are {\it trinions}, the
triples $(K, p_1, p_2, q)$, where $K$ is a Riemann surface of genus 0 with a
boundary $\partial K=\partial K^{\inp}_1\sqcup\partial
K^{\inp}_2\sqcup\partial K^{\outp}$ ($\partial K^{\inp}_i\simeq\partial
K^{\outp}\simeq\circle$), $p_i$ are input parametrizations of $\partial
K^{\inp}_i$ and $q$ is an output parametrization of $\partial K^{\outp}$. The
three copies of the Neretin semigroup $\Ner$ act on
$\Trinion(\Diff_+(\circle))$; the corresponding infinitesimal action of
$\CVect(\circle)+\CVect(\circle)+\CVect(\circle)$ on
$\Trinion(\Diff_+(\circle))$ is transitive. Let $\pi$ be the projective
representation of $\Ner\times\Ner\times\Ner$ in the space $\AP(V)$, the
$(\Ner\times\Ner\times\Ner)$--equivariant (or infinitesimally
$(\CVect(\circle)+\CVect(\circle)+\CVect(\circle))$--equivariant) mapping of
$\Trinion(\Diff_+(\circle))$ into $\AP(V)$ is called the projective
representation of the manifold of trinions $\Trinion(\Diff_+(\circle))$. The
next fact (which should be attributed to folklore, I think, and which is
convenient to formulate as a proposition) plays a crucial role.

\proclaim{Proposition 1} The manifold $\Trinion(\Diff_+(\circle))$ of trinions
for the group $\Diff_+(\circle)$ of diffeomorphisms of a circle admits one and
only one projective representation in the space $\Hom(V_{h_1,c}\otimes
V_{h_2,c},V_{h_3,c})$ for each central charge $c$ and for each triple of
weights $(h_1, h_2, h_3)$.
\endproclaim

The manifold of trinions $\Trinion(\Diff_+(\circle))$ will be also denoted
by\linebreak $\Trinion^+(\Diff_+(\circle))$, whereas
$\Trinion^-(\Diff_+(\circle))$ will denote the space of {\it anti-trinions},
the triples $(K, p, q_1, q_2)$, where $K$ is a Riemann surface of genus 0 with
a boundary $\partial K=\partial K^{\inp}\sqcup\partial
K^{\outp}_1\sqcup\partial K^{\outp}_2$ ($\partial K^{\inp}\simeq\partial
K^{\outp}_i\simeq\circle$), $p$ is an input parametrization of $\partial
K^{\inp}$ and $q_i$ are output parametrizations of $\partial K^{\outp}_i$. The
three copies of the Neretin semigroup $\Ner$ act on
$\Trinion^-(\Diff_+(\circle))$ and the corresponding infinitesimal action of
$(\CVect(\circle)+\CVect(\circle)+\CVect(\circle))$ on
$\Trinion^-(\Diff_+(\circle))$ is transitive.  The definition of the
projective representation of the manifold\linebreak
$\Trinion^-(\Diff_+(\circle))$ is the same as for
$\Trinion^+(\Diff_+(\circle))$. The following analog of Proposition 1
holds:{\sl the manifold $\Trinion^-(\Diff_+(\circle))$ of anti-trinions for
the group $\Diff_+(\circle)$ of diffeomorphisms of a circle admits one and
only one projective representation in the space
$\Hom(V_{h_1,c},V_{h_2,c}\otimes V_{h_3,c})$ for each central charge $c$ and
for each triple of weights $(h_1, h_2, h_3)$}.

It should be mentioned that trinions maybe glued with each other. On this way
we receive a structure, which underlies self--interacting string field theory
on tree level. Namely, the points of the manifold of {\it polynions\/}
$\Polynion(\Diff_+(\circle))$ are data $(K,p_1,\ldots,p_n,q)$, where $K$ is a
Riemann surface of genus 0 with a boundary $\partial
K=\bigsqcup_{i=1}^n\partial K^{\inp}_i\sqcup\partial K^{\outp}$ ($\partial
K^{\inp}_i\simeq\partial K^{\outp}\simeq\circle$), $p_i$ are input
parametrizations of $\partial K^{\inp}_i$ and $q$ is an output parametrization
of $\partial K^{\outp}$.  The manifold $\Polynion(\Diff_+(\circle))$ is
$\AZ_+$--graded, the connected component $\Polynion_n(\Diff_+(\circle))$ of
degree $n$ consists of data $(K,p_1,\ldots,p_{n-1},q)$; it should be mentioned
that $\Polynion_0(\Diff_+(\circle))=\Mantle(\Diff_+(\circle))$ and
$\Polynion_1(\Diff_+(\circle))=\Trinion(\Diff_+(\circle))$. The $(n+1)$ copies
of the Neretin semigroup $\Ner$ act on $\Polynion_n(\Diff_+(\circle))$ and the
corresponding infinitesimal action of $(n+1)\CVect(\circle)$ is transitive.
One may define a projective representation of the manifold
$\Polynion_n(\Diff_+(\circle))$ in the same way as for
$\Trinion(\Diff_+(\circle))$ and to formulate an analog of Proposition 1
(namely, that {\sl the manifold $\Polynion_n(\Diff_+(\circle))$ of polynions
for the group $\Diff_+(\circle)$ of diffeomorphisms of a circle admits one and
only one projective representation in the space $\Hom(\bigotimes_{i=1}^n
V_{h_i,c},V_{h_{n+1},c})$}). But the whole space $\Polynion(\Diff_+(\circle))$
admits a subsidiary structure defined by gluing. Namely, two elements
$(K^{(1)},p^{(1)}_1,\ldots p^{(1)}_n,q^{(1)})$ and $(K^{(2)},p^{(2)}_1,\ldots
p^{(2)}_m,q^{(2)})$ maybe glued in $m$ different ways into the third element
$(K^{(3)},p^{(3)}_1,\ldots p^{(3)}_{n+m-1},q^{(3)})$ as follows
$K^{(3)}=K^{(1)}\bigsqcup_{q^{(1)}(e^{it})=p^{(2)}_i(e^{it})} K^{(2)}$
($i=1,\ldots m$), $p^{(3)}_j=p^{(2)}_j$ if $j=1,\ldots i-1$, $p^{(1)}_{j-i+1}$
if $j=m,\ldots i+n-1$ and $p^{(2)}_{j-n+1}$ if $j=i+n,\ldots m+n-1$,
$q^{(3)}=q^{(2)}$. It should be mentioned that the gluing is compatible with
the grading and that $\Polynion_n(\Diff_+(\circle))$ ($n>1$) are generated by
$\Trinion(\Diff_+(\circle))=\Polynion_1(\Diff_(\circle))$ via gluing. Now one
may define the projective representation of $\Polynion(\Diff_+(\circle))$ as
the set of representations of $\Polynion_n(\Diff_+(\circle))$ in
$\Hom(H^{\otimes(n+1)},H)$, which correspond the composition of operators to
the gluing.

Here some subsidiary comments are needed. First of all, the definition of
polynions may be formulated also for anti-polynions (the set of which maybe
denoted by $\Polynion^-(\Diff_+(\circle))$ whereas the set of polynions maybe
denoted also by\linebreak $\Polynion^+(\Diff_+(\circle))$). Second, the
operation of the
gluing maybe generalized to the partial operation of the sewing: which
correspond to two polynions their gluing but with the extracting (if it is
possible) the element of the mantle from the input (or output) of the gluing.
The partial operation of the sewing maybe extended to the more large manifold
than one of polynions (see f.e. [18]). Thirth, it should be mentioned that the
definition of representation of $\Trinion(\Diff_+(\circle))$ do not deal with
operators from the space $\Hom(\cdot,\cdot)$, the definition of the
representation of polynions maybe formulated in analogous manner, it allows,
for example, to consider arbitrary linear relations instead of linear
operators (it is effective in infinite dimensional case [11]), etc; but below
we shall work with unbounded linear operators and linear relations as with
ordinary operators to avoid the "second--order" technical details of
functional analysis hidden in our infinite--dimensional geometric picture.
Fourth, the important class of representations of
$\Polynion(\Diff_+(\circle))$ is formed by the permutation--invariant ones;
namely, the representation of $\Polynion(\Diff_+(\circle))$ is called
permutation--invariant, if it is equivariant with respect to the actions of
symmetric groups $\operatorname{\bold S}_n$ on the inputs of elements of
$\Polynion_{n-1}(\Diff_+(\circle))$ and the multiples in tensor products
$H^{\otimes n}$. This is a natural class of representations but not unique
interesting one, for instance the permutation--invariant representations maybe
generalized to braided ones (in which permutations of multiples in tensor
product $H^{\otimes n}$ should be accompanied by some transforms in them).

The definition of the space $\Polynion(\Diff_+(\circle))$ and its
representations maybe generalized to the notion of {\it the Kontsevich--Segal
category\/} $\Train(\Diff_+(\circle))$, the train of the group
$\Diff_+(\circle)$ of diffeomorphisms of a circle, and its representations
(modular functors) [15,19] (see also [12,13]). Objects $\CA,\CB,\CC,\ldots$ of
the Kontsevich--Segal category are ordered finite sets (which are represented
by disjoint ordered unions of circles); morphisms from $\Mor(\CA,\CB)$ are
data $(K,p_1,\ldots p_n,q_1,\ldots q_m)$, where $K$ is arbitary Riemann
surface with a boundary $\partial K=\bigsqcup_{i=1}^n\partial
K^{\inp}_i\sqcup\bigsqcup_{j=1}^m\partial K^{\outp}_j$ ($n=\#\CA$, $m=\#\CB$),
$p_i$ are output parametrizations of $\partial K^{\inp}_i$ and $q_j$ are
output parametrizations of $\partial K^{\outp}_j$. The $(n+m)$ copies of the
Neretin semigroup $\Ner$ act on $\Mor(\CA,\CB)$ ($n=\#\CA$, $m=\#\CB$) but the
infinitesimal action of $(n+m)\CVect(\circle)$ is not transitive;
nevertheless, the representations of $\Mor(\CA,\CB)$ maybe correctly defined
in the same way as for $\Polynion(\Diff_+(\circle))$. The scope of
representations of $\Mor(\CA,\CB)$ forms a representation of the whole
category $\Train(\Diff_+(\circle))$ (or {\it modular functor\/}) if (1) they
transform the composition of morphisms (the gluing) into the composition of
operators, (2) if $K=K_1\sqcup K_2$ then the representation operator of
morphism corresponding to $K$ is a tensor product of ones corresponding to
$K_i$.

Some remarks should be done. First, one may include $\varnothing$ into the
class of objects of $\Train(\Diff_+(\circle))$ (see f.e. [15,19] but we prefer
not to do it). Second, the permutation--invariant modular functors maybe
defined in a way similar to the described above one for polynions.

\subhead 1.2. Infinitesimal objects for representations of
$\Polynion^{\pm}(\Diff_+(\circle))$:\linebreak QCFT--operator algebras and
coalgebras. Infinitesimal objects for per\-mu\-tation--invariant
representations
of $\Polynion^{\pm}(\Diff_+(\circle))$: vertex operator algebras and
coalgebras. Infinitesimal objects for modular functors: QCFT--operator
crossing--algebras and vertex operator crossing--algebras \endsubhead
The infinitesimal counterparts of objects of par. 1.1. will be constructed
below following ideas of [14,App.3] (see also [18]).

First, let describe the operation of {\it the vertex insertion\/} into the
element of the $\Mantle(\Diff_+(\circle))$ (the Neretin semigroup $\Ner$).
Namely, let us consider an arbitrary projective representation of the manifold
$\Trinion^+(\Diff_+(\circle))$ in $\Hom(V_{h_1,c}\otimes V_{h_2,c},
V_{h_3,c})$; the operator corresponding to the trinion $(K,p_1,p_2,q)$ will be
denoted by $\TY_{(K,p_1,p_2,q)}(\cdot,\cdot)$ or $\TY_{(K,p_1,p_2,q)}(\cdot)$,
where the second argument should stand out of brackets. Let us consider an
arbitrary element $(K,p,q)$ of $\Mantle(\Diff_+(\circle))$ and a point $z$ on
$K$; the operator $\lim_{p'(e^{it})\to z}\TY_{(K,p',p,q)}(v_{h_1})$ will be
considered as a result of an insertion of the vertex $v_{h_1}$ into $(K,p,q)$
at point $z$, where $v_{h_1}$ is the highest vector in $V_{h_1}$ and limit is
considered up to a multiple (i.e. in the projective space); it will be denoted
by $\TY_{(K,p,q)}(v_{h_1}; z)$ (it should be mentioned that this operator is
defined up to a number multiple). Second, one may define {\it a vertex\/}
itself as $\lim_{K\to\circle;p(e^{it}),q(e^{it})\to e^{it}} \TY_{(K,p,q)}(v_h;
z)$, the vertex (which is defined up to a multiple) will be denoted by
$\TY(v_h;z)$. In this construction $z\in\circle$, but $\TY(v_h;z)$ is a
well--defined operator if $|z|\ll 1$ as a rule (but this circumstance should
be considered as "second order" one).

It should be mentioned that the change of $\Trinion^+(\Diff_+(\circle))$
on\linebreak $\Trinion^-(\Diff_+(\circle))$ leads to the notion of a
co--vertex and the co--vertex insertion. So the vertex will be also denoted by
$\TY^+(v;z)$ and the result of its insertion by $\TY^+_{(K,p,q)}(v;z)$ whereas
their co--counterparts will be denoted by $\TY^-(v;z)$ and
$\TY^-_{(K,p,q)}(v;z)$.

Now let us consider the situation, when a representation of
$\Trinion^{\pm}(\Diff_+(\circle))$ is extended to a representation of
$\Polynion^{\pm}(\Diff_+(\circle))$. It is natural that the gluing operation
in the least should induce some algebraic structure on vertices
(co--vertices). Indeed, this case vertices (co--vertices) form a closed
operator algebra (co--algebra) in the sense specified below. Some definitions
are necessary here [20-22].

\definition{Definition 1}

{\bf A.} {\it A QFT--operator algebra\/} ({\it operator algebra of quantum
field
theory\/}) is a pair $(H,t^{\gg}_{\ga\gb}(\vec x))$, where $H$ is a vector
space and $t^{\gg}_{\ga\gb}(\vec x)$ is a $H$--valued tensor field on $\AR^n$
or $\AC^n$ such that $t^{\gd}_{\ga\gz}(\vec x)t^{\gz}_{\gb\gg}(\vec
y)=t^{\gz}_{\ga\gb}(\vec x-\vec y)t^{\gd}_{\gz\gg}(\vec y)$.

{\bf B.} {\it A QFT--operator coalgebra\/} ({\it operator coalgebra of quantum
field theory\/}) is a pair $(H,t^{\ga\gb}_{\gg}(\vec x))$, where $H$ is a
vector space and $t^{\ga\gb}_{\gg}(\vec x)$ is a $H$--valued tensor field on
$\AR^n$ or $\AC^n$ such that $t^{\gb\gz}_{\ga}(\vec x)t^{\gg\gd}_{\gz}(\vec
y)=t^{\gz\gd}_{\ga}(\vec y)t^{\gb\gg}_{\gz}(\vec x-\vec y)$.

\enddefinition

It should be mentioned that if $H$ is a QFT--operator algebra then $H^*$ is
QFT--operator coalgebra and vice versa. So below we shall be interested
presumably in QFT--operator algebras.

Let us define the operators $l_{\vec x}(e_{\ga})$ ($l_{\vec
x}(e_{\ga})e_{\gb}=t^{\gg}_{\ga\gb}(\vec x)e_{\gg}$) in QFT--operator
algebras, then the following identities will hold $l_{\vec x}(e_{\ga})l_{\vec
y}(e_{\gb})=t^{\gg}_{\ga\gb}(\vec x-\vec y)l_{\vec y}(e_{\gg})$ ({\it operator
product expansion\/}) and $l_{\vec x}(a)l_{\vec y}(b)=l_{\vec y}(l_{\vec
x-\vec y}(a)b)$ ({\it duality\/}). On the other hand one may introduce the
multiplication operation $m_{\vec x}:H\otimes H\mapsto H$ as $m_{\vec
x}(a,b)=l_{\vec x}(a)b$, then $m_{\vec
x}(\id\otimes m_{\vec y})=m_{\vec y}(m_{\vec x-\vec y}\otimes\id)$.
Analogously, one may introduce the comultiplication operation $\gD_{\vec
x}:H\mapsto H\otimes H$ as $\gD_{\vec x}(e_{\ga})=t_{\ga}^{\gb\gg}(\vec
x)e_{\gb}\otimes e_{\gg}$ in the QFT--operator coalgebra, then
$(\id\otimes\gD_{\vec y})\gD_{\vec x}=(\gD_{\vec x-\vec y}\otimes\id)\gD_{\vec
y}$.

Below we shall need in a specific class of QFT--operator (co)algebras [20-22].

\definition{Definition 2} A QFT--operator algebra
$(H,t^{\gg}_{\ga\gb}(u);u\!\!\in\!\!\AC)$ is called {\it a QPFT--ope\-rator
algebra\/}
({\it operator algebra of the quantum projective field theory\/}) iff (1) $H$
is a direct sum of Verma modules $V^a$ over Lie algebra $\sLtwo$ with highest
vectors $v^a$ of highest weights $h^a$, (2) the operator fields $l_u(v^a)$ are
$\sLtwo$--primary (quasi--primary in terminology of [23]) of weight $h^a$:
$[L_k,l_u(v^a)]=(-u)^k(u\frac d{du}+(k+1)h^a)l_u(v^a)$
($[L_i,L_j]=(i-j)L_{i+j}$), (3) the rule of descendant generation holds:
$l_u(L_{-1}\Phi)=L_{-1}l_u(\Phi)$. A QFT--operator algebra is called {\it a
derived QPFT--operator algebra\/} if conditions (1),(2) and the derived rule
of descendant generation ($l_u(L_{-1}\Phi)=[L_{-1},l_u(\Phi)]=\frac
d{du}l_u(\Phi)$) hold.
\enddefinition

As it was shown in [21] the categories of QPFT--operator algebras and derived
QPFT--operator algebras are equivalent. The equivalency is realised in a
rather simple manner described in [21]. Below we shall not distinct both types
of algebras considering them as two faces of one object.

\definition{Definition 3}

{\bf A.} A highest vector $T$ in the QPFT--operator algebra is called {\it the
conformal stress--energy tensor\/} iff the operator field $T(u)=l_u(T)$ has an
expansion $T(u)=\sum_{k\in\AZ}L_k u^{-k-2}$, where $L_k$ form the Virasoro
algebra $\Cvir$: $[L_i,L_j]=(i-j)L_{i+j}+\gd(i+j)\frac{j^3-j}{12}\cdot 1$.

{\bf B.} A QPFT--operator algebra with fixed conformal stress--energy tensor
is called {\it a QCFT--operator algebra\/} ({\it operator algebra of the
quantum
conformal field theory\/}) {\it in a wide sense}.

{\bf C.} A QCFT--operator algebra in a wide sense is called {\it a
QCFT--operator algebra in a narrow sense\/} iff its space $H$ is a sum of the
Virasoro highest weight modules $V^a$, which highest vectors $v^a$ are
$\Cvir$--primary (i.e. the identity (2) of def.2 holds for all generators of
$\Cvir$).
\enddefinition

It should be mentioned that an arbitrary QCFT--operator algebra in a narrow
sense admits a strict representation by matrices with coefficient in the
algebra $\Vrt(\Cvir;c)$ of vertex operators for the Virasoro algebra [20].

It should be mentioned that an analog of Def.2 may be formulated for
QPFT--coalgebras. Also one should mention that def.3C may be formulated
without supposition that generators of the Virasoro algebra form the conformal
stress--energy tensor (see f.e.[20]), so one may define QCFT--operator
coalgebra in such manner.

Let us now define several main objects of this paragraph.

\definition{Definition 4}

{\bf A.} {\it A crossing--algebra\/} is a triple $(H,m,\gD)$, where the
mappings $m:H\otimes H\mapsto H$ and $\gD:H\mapsto H\otimes H$ define
structures of associative algebra and associative coalgebra on $H$ such that
$(\id\otimes m)(\gD\otimes\id)=\gD m=(m\otimes\id)(\id\otimes\gD)$, such
identity should also hold after a change of $m$ on $m'$ ($m'(a,b)=m(b,a)$).

{\bf B.} {\it A QFT--operator crossing--algebra\/} is the triple $(H,m_{\vec
x},\gD_{\vec x})$, where the mappings $m_{\vec x}:H\otimes H\mapsto H$ and
$\gD_{\vec x}:H\mapsto H\otimes H$ define structures of QPFT--operator algebra
and coalgebra on $H$ such that $\gD_{\vec x-\vec y}m_{\vec y}=(\id\otimes
m_{\vec y})(\gD_{\vec x}\otimes\id)$, $\gD_{\vec x}m_{\vec x-\vec
y}=(\id\otimes m_{\vec y})(\gD_{\vec x}\otimes\id)$, such identity should also
hold after a change of $m$ on $m'$ ($m'(a,b)=m(b,a)$).

{\bf C.} {\it A QPFT--operator crossing--algebra\/} is a QFT--operator
crossing--algebra built from QPFT--operator algebra and QPFT--operator
coalgebra; {\it a QCFT--operator crossing--algebra\/} is a QPFT--operator
crossing algebra built from QCFT--operator algebra and QCFT--operator
coalgebra.
\enddefinition

\proclaim{Remark} A crossing algebra $H$ supplied by commutator operations is
a Lie bialgebra.
\endproclaim

The first main proposition of this paragraph may be formulated as follows.

\proclaim{Proposition 2}

{\bf A.} The space $H$ of any representation of the manifold
$\Polynion^+(\Diff_+(\circle))$ is supplied by a structure of a QCFT--operator
algebra.

{\bf B.} The space $H$ of any representation of the manifold
$\Polynion^-(\Diff_+(\circle))$ is supplied by a structure of a QCFT--operator
coalgebra.

{\bf C.} The space $H$ of any representation of the category
$\Train(\Diff_+(\circle))$, the train of the group $\Diff_+(\circle))$ of
diffeomorphisms of a circle, is supplied by a structure of QCFT--operator
crossing--algebra (in a narrow sense).

\endproclaim

Namely, the vertices and co--vertices generate the structure of such algebra.
This fact on a tree level should be attributed to folklore.
The compatibility conditions for the multiplication and the co--multiplication
in the QCFT--operator crossing--algebra is a natural sequence of the
cutting--gluing conditions for representations of $\Train(\Diff_+(\circle))$.

\definition{Definition 5}
QCFT--operator algebra is called {\it a vertex operator algebra\/} iff
operator fields $l_u(\Phi)$ are mutually local i.e. $[l_u(\Phi),l_v(\Psi)]=0$
if $u\ne v$.
\enddefinition

The more formal definition maybe found in [24,23,17] (there are claimed also
that (1) all operator product expansions are meromorphic, (2) the weights of
all elements are integral, (3) the spaces of a fixed weight are
finite--dimensional and empty for a sufficiently small weights, but for our
purposes these conditions are excessive). The locality condition maybe
rewritten as $m_u(m_v\otimes\id)=m_v(m_u\otimes\id)$ if $u\ne v$. So one may
define {\it a vertex operator coalgebra\/} as QCFT--operator coalgebra such as
$(\gD_u\otimes\id)\gD_v=\gD_v(\gD_u\otimes\id)$ and {\it a vertex operator
crossing--algebra\/} as a QCFT--operator crossing--algebra, which as
QCFT--operator (co)algebra is a vertex operator (co)algebra.

The second main proposition of this paragraph maybe formulated as follows.

\proclaim{Proposition 3}

{\bf A.} The space $H$ of any permutation--invariant representation of the
manifold $\Polynion^+(\Diff_+(\circle))$ is supplied by a structure of a
vertex operator algebra.

{\bf B.} The space $H$ of any permutation--invariant representation of the
manifold $\Polynion^-(\Diff_+(\circle))$ is supplied by a structure of a
vertex operator coalgebra.

{\bf C.} The space $H$ of any permutation--invariant representation of the
category
$\Train(\Diff_+(\circle))$, the train of the group $\Diff_+(\circle))$ of
diffeomorphisms of a circle, is supplied by a structure of vertex operator
crossing--algebra.

\endproclaim

It is partially (on a tree level) contained in [18] (in a slight different
language based rather on the concept of the sewing than on one of the gluing).

Some remarks are necessary. First, the vertex operator algebras were recently
described in the new--fashion operadic terms [25], it will be very interesting
to give analogous interpretations for other objects of this paragraph. Second,
it is rather interesting to investigate finite--dimensional ordinary (not
operator) crossing--algebras --- the simplest examples related to matrix
algebras and group rings for finite groups seem to be very interesting. Third,
the generalization of the representation theory of vertex operator algebras
[26] on other objects of the paragraph maybe rather interesting. Fourth, one
may extend the category $\Train(\Diff_+(\circle))$ by supplying its morphisms
by additional structures (f.e. by the fixed polarization in (co)homologies,
see [13]), in this case the crossing--identities are broken and it is
interesting to explicate their reduced form. Fifth, one mey generalize Def.5
on arbitrary QFT--operator algebras to obtain the object called {\it vertex
algebra\/} (see f.e. [18]), similarly one may define {\it vertex coalgebras\/}
and {\it vertex crossing--algebras}. Sixth, it should be mentioned that in
papers [13,20,21] the term "vertex operator algebra" is used in the sense of
"QFT--operator algebra". Seventh, one may define {\it vertex superalgebras\/}
and {\it vertex operator superalgebras\/} if changes the claim of locality
(commutativity of operators $l_{\vec x}$) on its superanalog.

\subhead 1.3. Renormalization of pointwise product in QCFT--operator algebras:
local conformal field algebras (LCFA). The renormalized Krichever--Novikov
functor \endsubhead Let us now imbed our constructions into a framework of the
noncommutative geometry following [27,4]. First, we shall formally describe
the result of the renormalization of pointwise product in QCFT--operator
algebras; second, we shall indicate a concrete renormalization procedure. It
should be mentioned that all our constructions maybe considered as based on
the initial idea of E.Witten [28] adapted to QCFT (for some hints see f.e.
[29]).

In general, the approach is based on the following considerations [4]. The
classical conformal field is a tensorial quantity on a complex curve; the set
of such fields can be regarded as a quantity on a covering of this curve, and
quantization means the transition to noncommutative coverings. Every curve has
inner degrees of freedom (modules) and therefore has to de considered with all
its deformations. It is natural to assume that a deformation of a
noncommutative object is noncommutative, and this means that we heve deal with
quantum deformations --- noncommutative spaces "projected" onto the
noncommutative bases. The separate fibers cannot be isolated from each other
and so have to be considered collectively. Thus, we do not deal with
individual fields but with families of conformal quantum fields. Below the
curve is a unit complex disc.

Some constructions are necessary [4,7]. {\it A model of the Verma modules over
the Virasoro algebra\/} $\Cvir$ is a direct integral of the Verma modules
$V_{h,c}$ with the fixed central charge $c$ over this algebra. The simplest
realisation of the model is that of I.N.Berbshtein, I.M.Gelfand and
S.I.Gelfand: the model space is the Fock space over the fundamental affine
space for $\widetilde{\Vir}$, the universal covering of the fundamental affine
space $A(\Vir)$ for $\Vir$. The fundamental affine space $A(\Vir)$ is
stratified over the flag manifold $M(\Vir)$ with fibre $\AC^*$, so the model
space can be identified with the space of analytic functions of an infinite
set of variables $t, c_1, c_2, c_3, \ldots, c_n,\ldots$ (there are admitted
arbitrary degrees of $t$), the action of the Virasoro algebra in the model is
defined by the following formulas $L_{-p}=c_p+\sum_{\sssize{k\ge
1}\tsize}c_{k+p}\sdp{c_k}$ ($p>0$), $L_0=\sum_{\sssize{k\ge 1}\tsize}
kc_k\sdp{c_k}-t\sdp{t}$, $L_1=\sum_{\sssize{k\ge 1}\tsize}
c_k((k+2)\sdp{c_{k+1}}\!-\!2\sdp{c_1}\sdp{c_k})-t\sdp{t} \sdp{c_1}$,
$L_2=\sum_{\sssize{k\ge 1}\tsize}
c_k((k+3)\sdp{c_{k+2}}\!-\!(4\sdp{c_2}\!-\!(\sdp{c_1})^2)\sdp{c_k}\!-\!
b_k(\sdp{c_1},\ldots\sdp{c_{k+2}}))-t\sdp{t}(4\sdp{c_2}\!-\!(\sdp{c_1})^2)+
{\sssize{\frac {c}2}}(\sdp{c_2}\!-\!(\sdp{c_1})^2)$,
$L_n={\sssize{\frac{(-1)^n}{(n-2)!}}}\ad^{n-2}L_1\cdot L_2$ ($n>2$); the
highest vector $v_h$ of weight $h$ has the form $t^{-h}$. Another realisation
of the model is one in the Fock space over the universal deformation of a
complex disc [7]. Namely, if $f$ is the function from class $S$ then the
function $f(1-wf)^{-1}$ has the same property iff $w^{-1}$ is not contained in
the image of $f$. The set of such points forms a doamin
$\AC\backslash(f(D_+))^{-1}$ in the complex plane $\AC$. The union of the
pairs $(f,w)$, where $f\!\in\!S$ and $w^{-1}\nin f(D_+)$ is the space
$A(\Vir)$ of the universal deformation of a complex disc; the map
$(f,w)\mapsto f$ is the projection onto the basis, which coincides with the
flag space $M(\Vir)$ for the Virasoro--Bott group $\Vir$. The model space is
the same as in the BGG--realisation; the action of the Virasoro algebra is
given by the formulas $L_{-p}=c_p+\sum_{\sssize{k\ge
1}\tsize}c_{k+p}\sdp{c_k}$ ($p>0$), $L_0=\sum_{\sssize{k\ge 1}\tsize}
kc_k\sdp{c_k}-t\sdp{t}$, $L_1=\sum_{\sssize{k\ge 1}\tsize}
c_k((k+2)\sdp{c_{k+1}}\!-\!2\sdp{c_1}\sdp{c_k})-t\sdp{t} \sdp{c_1}$,
$L_2=\sum_{\sssize{k\ge 1}\tsize}
c_k((k+3)\sdp{c_{k+2}}\!-\!(4\sdp{c_2}\!-\!(\sdp{c_1})^2)\sdp{c_k}\!-\!
b_k(\sdp{c_1},
\ldots\sdp{c_{k+2}}))-t\sdp{t}(4\sdp{c_2}\!-\!(\sdp{c_1})^2)+t^2\sdp{t}+
{\sssize{\frac {c}2}}(\sdp{c_2}\!-\!(\sdp{c_1})^2)$,
$L_n={\sssize{\frac{(-1)^n}{(n-2)!}}}\ad^{n-2}L_1\cdot L_2$ ($n>2$); the
highest vector $v_h$ of the weight $h$ has the form $t^{-h}G(h,c; tc_1,
t^2c_2, t^3c_3,\ldots t^kc_k,\ldots)$, where the function $G(h,c; u_1, u_2,
u_3, \ldots u_k, \ldots)$ of the infinite set of variables $u_1, u_2,
u_3,\ldots u_k,\ldots$ satisfies a certain system of differential equations
[7,27], which allows to consider it as the confluent hypergeometric function
in the sense of Gelfand et al, corresponding to the double fibration of the
universal deformation of the complex disc, which bases are the complex disc
and the flag space of the Virasoro--Bott group.

It turns out that the model of the Verma modules over the Lie algebra $\Cvir$
is equipped with a richer structure than manifested one [27,4].

Let's remind that if $\CR$ is an associative algebra with identity over a field
$\Bbbk$ and $\Fg$ is the Lie subalgebra of the algebra $\Der(\CR)$ nof
derivations
of $R$ then an associative algebra $A$ with identity over $\Bbbk$ is called an
{\it L--algebra over the pair\/} $(\CR,\Fg)$ iff $\CA$ is right $\CR$--module,
the
mapping $r\mapsto r\cdot 1$ from $\CR$ to $\CA$ is a ring morphism, (and,
therefore,
$\CA$ is $\CR$--bimodule), $\CA$ is a $\Fg$--module such that the $\Fg$--module
structure is compatible with the left $\CR$--module structure.

\define\reg{\operatorname{reg}}
\define\spn{\operatorname{span}}
Let $\CO=\CO(\tilde{\AC}^*)$ and $\Cvir_{\reg}=\spn(L_p, p=-1,0,1,2,\ldots)$.
The model of the Verma modules over the Virasoro algebra, realized in the Fock
space over the universal deformation of a complex disc, admits a natural right
$\CO$--module structure via multiplication by functions of a single variable
$t$. As it was shown in [27] the model of the Verma modules over the Lie
algebra $\Cvir$, realized in the Fock space over the universal deformation of
the complex disc, possesses exactly one structure of L--algebra over
$(\CO,\Cvir_{\reg})$, compatible with the right $\CO$--module structure in the
model, such that $L_pT(\Phi)=T(L_p\Phi)$ ($p=-1,-2,-3,-4,\ldots$; $T$ is the
operator of the left multiplication in the L--algebra) for every element
$\Phi$ of the model. The operator $T$ is defined as $T(c_p)=L_{-p}$,
$T(t)=\sum_{k\ge 0}t^{1-k}P_k(\sdp{c_1},\ldots,\sdp{c_k})\cdot (-1)^k$, where
the polynomials $P_k$ have the form $P_0=0$, $P_1(u_1)=u_1$, $P_2(u_1,
u_2)=u_2-u_1^2$, $P_3(u_1, u_2, u_3)=u_3-3u_1u_2+2u^3_1$, $P_4(u_1, u_2, u_3,
u_4)=u_4-4u_1u_3+10u_1^2u_2-5u_1^4-2u_2^2$, $P_{k+1}=(k+2)^{-1}(2(1-k)u_1P_k+
\sum_{j\ge 1}((j+2)u_{j+1}-2u_1u_j)\sdp{u_j}P_k-\sum_{i,j>1; i+j=k+1}
P_iP_j)$. The resulting algebra is denoted by $L(\Cvir; c)$. A careful analysis
of the above construction enables us to interpret the L--algebra
$L(\Cvir)$ as a structure ring of the quantum universal deformation of a
complex disc [4].

The representation of elements of $L(\Cvir)$ by elements of the model of the
Verma modules over $\Cvir$, realized in the Fock space over the universal
deformation of a complex disc, is called {\it a direct recording}. The direct
recording of an element $\Phi$ will be denoted by $[\Phi]_d$. Let's also
introduce {\it a reverse recording\/} [27,4]: we say that
$t_{\lambda}c_1^{k_1}\ldots c_n^{k_n}$ is the reverse recording of the element
$\Phi$ of the L-algebra $L(\Cvir)$ and we write $[\Phi]_r=t^\lambda
c^{k_1}_1\ldots c_n^{k_n}$ if $\Phi=T(t)^\lambda c_1^{k_1}\ldots c_n^{k_n}$.
As it was shown in [27] the action of the Virasoro algebra $\Cvir$ on the
L-algebra $L(\Cvir)$ with the reverse recording of elements has the form
$L_{-p}=c_p+\sum_{\sssize{k\ge 1}\tsize}c_{k+p}\sdp{c_k}+(-t)^{1-p}\sdp{t}$
($p>0$), $L_0=\sum_{\sssize{k\ge 1}\tsize}kc_k\sdp{c_k}-t\sdp{t}$,
$L_1=\sum_{\sssize{k\ge 1}\tsize}c_k((k+2)\sdp{c_{k+1}}\!-
\!2\sdp{c_1}\sdp{c_k})+t^2\sdp{t}$, $L_2=\sum_{\sssize{k\ge 1}\tsize}
c_k((k+3)\sdp{c_{k+2}}\!-\!(4\sdp{c_2}\!-\!(\sdp{c_1})^2)\sdp{c_k}\!-\!
b_k(\sdp{c_1},\ldots\sdp{c_{k+2}}))+ {\sssize{\frac
{c}2}}(\sdp{c_2}\!-\!(\sdp{c_1})^2)-t^3\sdp{t}$,
$L_n={\sssize{\frac{(-1)^n}{(n-2)!}}}\ad^{n-2}L_1\cdot L_2$ ($n>2$).

Several aditional definitions are needed [27,4].
Namely, an L-algebra $\CB$ over the pair $(\CR,\Fg)$ is called {\it an
L$^\circ$-algebra over\/} $(\CR,\Fg)$ iff $\Fg\subseteq\Der(\CB)$. Let $\CA$ be
an
L-algebra over $(\CR,\Fg)$ and $\CB$ be an L$^\circ$-algebra over the same
pair.
An L-algebra $\CC$ over $(\CR,\Fg)$ is called {\it a local field algebra with
the
algebra of primary fields $\CB$ and the structure algebra $\CA$\/} if
(1) $\CC$ is a left $\CA$--module, (2) $\CC$ is an L-algebra over the pair
$(\CB,\Fg)$.
If $\CR=\CO$, $\Fg=\Cvir_{\reg}$ and $A=L(\Cvir)$ then $\CC$ is called {\it a
local
conformal field algebra\/} ({\it LCFA\/}).
The local conformal field algebras maybe regarded as structure rings of the
quantum universal deformations of noncommutative coverings of a complex disc
according to [4]. It should be marked that the analogs of direct and reverse
recordings maybe defined ifor an arbitrary LCFA. It allows to define {\it a
canonical connection\/} in an LCFA $\CC$, namely, if $\CC$ is identified with
$\AC[c_1, c_2, \ldots c_n, \ldots]\otimes\CB$ via the reverse recording then
$\nabla=1\otimes l_{-1}$, where $l_{-1}$ is the action of the element $e_{-1}$
of the Virasoro algebra $\Cvir$ in $\CB$.

On the other hand, LCFAs can be regarded as the result of the renormalization
of a pointwise product of fields in the corresponding field theories. Let's
describe the concrete renormalization procedure.

Namely, let $H$ be an arbitrary QCFT--operator algebra. Let's denote
$\CO\otimes H$ by $\CA(H)$. The structure of the LCFA in $\CA(H)$ is defined
as follows. Let $\varphi$ be an arbitrary element of $H$, let's define
$\varphi(f)$ as $\res\{f(u)l_u(\varphi)\frac{du}u\}=\lim_{u\to
0}\{f(u)l_u(\varphi)-\text{ singularities }\}$ ($f\!\in\!\CO$);
$\varphi(f)$ maybe considered as an operator of the left multiplication on
$(f,\varphi)$ in $\CA(H)$. The Virasoro algebra $\Cvir$ naturally acts in
$\CA(H)$. All these data determine a structure of an LCFA in $\CA(H)$.

Below some additioonal structures on the LCFAs will be needed. They are point
projectors $P_t:\CA(H)\mapsto H$ and point imbeddings $I_t:H\mapsto\CA(H)$.
The point projector $P_t$ correspond the element $f(t)\varphi$ to the pair
$(f,\varphi)$. The point imbedding is its right inverse, i.e. $P_tI_t=\id$.
The additional condition, which determines the point imbedding $I_t$ is that
$\nabla I_t=0$, where $\nabla$ is the canonical connection in the LCFA.

Let's very shortly describe a renormalized analog of the Krichever--Novikov
construction of operator product expansions on Riemann surfaces [30] combining
it simultaneously with the Kontsevich--Segal approach. If the initial
QCFT--operator algebra is meromorphic (i.e. all operators $l_u(\varphi)$ are
meromorphic) then one may construct a sheaf $\CA_{\Xi}$ on an Riemann surface
$R_{\Xi}$ (representing a morphism $\Xi$ from $\Train(\Diff_+(\circle))$) from
the algebra $\CA(H)$. So one corresponds to a Riemann surface $R_{\Xi}$ its
noncommutative covering, which structural ring consists of global sections of
$\CA_{\Xi}$ over $R_{\Xi}$. This correspondence maybe considered as {\it a
renormalized Krichever--Novikov functor}. This construction maybe generalized
on QCFT--operator algebras with half--integer spectrum of primary fields by
the supplying the Riemann surfaces by a spinor structure. One may also
consider a general case with some minor neglegence of the full mathematical
rigor.

\head 2. Infinite dimensional geometry and topological conformal field
theory
\endhead

One of the main features of a consistent string field theory, which extracts
it from all 2D QCFT, is that it is a topological conformal field theory
(TCFT), i.e. roughly speaking, it contains a "correctly" defined
BRST--operator; these chapter is devoted to an adaptation of our infinite
dimensional geometric picture to such theories --- it should be mentioned that
from the point of view of the noncommutative geometry topological conformal
field theories are naturally appeared in the framework of quantum conformal
field theories as their "noncommutative de Rahm complexes" (cf. [10,11]).
There exist several approaches to TCFT and they are, in general, similar to
ones described in the first chapter. But now a reverse order of a presentation
is more preferable: BRST--operators and formal cohomological machinery in a
general framework of the quantum projective field theory will be described
following [31,par.2.1] in par.2.1; these concepts are adapted to QCFT in
par.2.2.; at least, the infinitesimal picture is "exponentiated" to the
concept of a string background in par.2.3.

\subhead 2.1. Differential and topological QFT--operator algebras and their
cohomology; topological QPFT--operator algebras; stress--energy tensors,
currents and their charges, BRST--currents and ghost fields in QPFT--operator
algebras
\endsubhead
Below we shall consider QFT--operator algebras with identity (i.e. an element
$1$ such that $l_{\vec x}(1)=\id$). There is defined a vector--operator $\vec
L$ as $\vec L\Phi=\left.\frac{d}{d\vec
x}l_{\vec x}(\Phi)1\right|_{\vec x=0}$ . It belongs to $\Der(\FA)$, i.e. $[\vec
L,l_{\vec
x}(\Phi)]=l_{\vec x}(\vec L\Phi)$.

\definition{Definition 6}

{\bf A.} The $\AZ$--graded QFT--operator algebra $\FA$ is called {\it a
differential QFT--operator algebra\/} if there is defined an element $Q$
(called {\it a BRST--operator\/}) of degree 1 in $\Der(\FA)$ such that
$Q^2=0$.

{\bf B.} The differential QFT--operator algebra $\FA$ is called {\it a
topological QFT--operator algebra\/} if there exist a vector--operator $\vec B$
such that
$[Q,\vec B]=\vec L$ (here and below $[\cdot,\cdot]$ is a supercommutator).

{\bf C.} The QFT--operator algebra is called topological vertex superalgebra
if it is a topological QFT--operator algebra and vertex superalgebra
simultaneously.

\enddefinition

Topological vertex superalgebras were considered in [32].
Now let's formulate the first main proposition
of this paragraph.

\proclaim{Proposition 4}

{\bf A.} The cohomology $H^*(\FA)$ of a differential QFT--operator algebra
forms a QFT--operator algebra.

{\bf B.} The cohomology $H^*(\FA)$ of a topological QFT-operator algebra forms
an associative algebra.

{\bf C.} The cohomology $H^*(\FA)$ of a topological vertex superalgebra forms
an associative supercommutative algebra.

\endproclaim

The demonstrations from [33] maybe straightforwardly adapted to our case.

Let's now consider the case of QPFT--operator algebras. {\it Differential
QPFT--operator algebras\/} maybe defined as QPFT--operator algebras, which are
differential QFT--operator algebras. If there exist an operator $B$ in the
differential QPFT--operator algebra such that $L_{-1}=[Q,B]$ (it should be
mentioned that in the derived QPFT--operator algebras $L=L_{-1}$) than the
$\sLtwo$--module generated from $B$ by an adjoint action of $Q$ admits an
epimorphism onto the adjoint module. We shall claim that this epimorphism maybe
split, i.e. one may find three operators $B_{-i}$ ($i=-1,0,1$; $B_{-1}=B$),
which transforms under $\sLtwo$ as elements of an adjoint module, i.e.
$[L_i,B_j]=(i-j)B_{i+j}$. Differential QPFT--operator algebra with additional
data ($Q$, $B_i$) will be called {\it topological QPFT--operator algebra}. It
should be mentioned that we do not know the commutation relations of $B_i$;
they should be determined in each concrete case, so one may say that $B_i$
generate hidden symmetries, which will be called {\it hidden ghost
symmetries}. It is a rather important problem to set them free in the
concrete case (cf. [34]).

Let's formulate the second main proposition of this paragraph.

\proclaim{Proposition 5} Operator $B_0$ in the topological QPFT--operator
algbera $\frak A$ defines an operator $\hat B_0$ in its cohomology $H^*(\frak
A)$ such that $\hat B^2_0=0$. If $\frak A$ is also a topological vertex
superalgebra then $H^*(\frak A)$ is a Batalin--Vilkovisky algebra.
\endproclaim

The proof of this fact maybe obtained by a straightforward adaptation of
arguments from [33]. One should see also [35] for a definition of {\it
Batalin--Vilkovisky algebras\/} and [32] for their operadic formulations.

Some remarks are convenient. First, it is very interesting to apply the
described cohomological machinery to the concrete QPFT --- the
$q_R$--conformal and super--$q_R$--conformal field theories [31,par.2.2,2.3].
Second, Prop.5 has its co-- and crossing counterparts; nevertheless, the
corresponding co-- and crossing analogs of Batalin--Vilkovisky algebras were
not investigated. Third, it is rather interesting, to unravel a
Batalin--Vilkovisky--like structure in a general case of topological
(non--local) QPFT--operator algebras.

Now let us call the meromorphic primary operator field of weight $1$
in the QPFT--operator algebra by {\it a current\/} (see [31,par.2.1]). Each
current has {\it charge\/} -- the coefficient in the Laurent expansion by
$u^{-1}$; charges of currents form a Lie algebra [31,par.2.1].
Let us call the primary operator field of weight $2$ by {\it a stress--energy
tensor\/} iff its three components form the Lie algebra $\sLtwo$; it should be
mentioned that in an arbitrary QPFT--operator algebra the stress--energy
tensor is not obligatory local (see f.e. [31,par.2.2] for $q_R$--conformal
field theories as examples).

\proclaim{Remark}

{\bf A.} In each QPFT--operator algebra the generators of the Lie algebra
$\sLtwo$ uniquely define a stress--energy tensor.

{\bf B.} In each differential QPFT--operator algebra the BRST--operator
uniquely defines a corresponding current ({\it a BRST--current\/}).

{\bf C.} In each topological QPFT--operator algebra the generators of the
hidden ghost symmetries uniquely define a primary operator field of weight 2
({\it ghost field\/}).

\endproclaim

These statements are sequences of the main structural theorem for
QPFT--operator algebras identifying them with subalgebras of
$\Mat_n(\Vrt(\sLtwo))$, where $\Vrt(\sLtwo)$ is a special QPFT--operator
algebra --- the algebra of vertex operators for the Lie algebra $\sLtwo$
[31,22] and the explicit formulas for primary fields in $\Vrt(\sLtwo)$ [36].

In general, neither stress--energy tensor nor a BRST--current or a ghost field
do not correspond to any elements of a QPFT--operator algebra. Nevertheless,
we can add them so below it will be supposed that stress-energy tensor,
BRST--current and ghost field belong to the considered QPFT--operator algebra.

\subhead 2.2. Topological conformal field theories and $N\!=\!2$ superconformal
field
theories
\endsubhead
Let's now adapted the general picture of par.2.1. to quantum conformal field
theories. In this case the stress--energy tensor is local and it seems to be
natural to claim the BRST--current and the ghost field to be local, too.
We also suppose that there exist {\it the ghost number counting current\/}
$J(z)$. All such claims allow to transform the topological conformal field
theory into the $N\!=\!2$ superconformal one by the twist $L(z)\to
L(z)+\frac12\partial J(z)$ (here $L(z)$ is a stress--energy tensor). The ghost
field and the BRST--current are transformed into the fermionic fields
$G^{\pm}(z)$ of spin $\frac32$ such that the commutation relations of the
components hold:
$$\allowdisplaybreaks
\align
[L_n,L_m]=&(n-m)L_{n+m}+\frac14d(n^3-n)\delta(n+m)\\
[L_n,G^{\pm}_m]=&(\frac12 n-m)G^{\pm}_{n+m}\\
[L_n,J_m]=&-mJ_{m+n}\\
[G^+_n,G^-_m]=&L_{n+m}+\frac12(n-m)J_{n+m}+\frac12d(n^2-\frac14)\delta(n+m)\\
[J_n,J_m]=&dn\delta(n+m)\\
[J_n,G^{\pm}_n]=&\pm G^{\pm}_{n+m}\\
[G^{\pm}_n,G^{\pm}_m]=&0.
\endalign
$$
It should be also mentioned that topological conformal field theories (or
equivalently $N\!=\!2$ superconformal ones) possess additional algebraic
structures related to so--called {\it homotopy Lie algebras\/} (see [37-39]).

Let's consider the renormalization of pointwise product of fields in the
QCFT--operator algebras of the topological conformal field theory. Let
$\CA=\CO(\tilde{\AC}^*,H)$ be the corresponding LCFA.

\proclaim{Proposition 6} The LCFA $\CA=\CO(\tilde{\AC}^*,H)$ ($H$ is the
space of the operator algebra of the topological conformal field theory) maybe
enlarged to the complex $\CA^{\cdot}=\Omega^{\cdot}(\tilde{\AC}^*,H)$
with the differential $D=d+Q$ ({\it an enlarged BRST--operator\/}), where $d$
is a natural differential in $\Omega^{\cdot}(\tilde{\AC}^*)$ and $Q$ is the
BRST--operator in $H$.
\endproclaim

\demo{Proof} One should introduce a new variable $dt$ and postulate the
commutation relations between $dt$ and $\xi\!\in\!\CA$ of the form:
$[dt,\xi]=D([t,\xi])-[t,Q\xi]$.
\enddemo

$\CA^{\cdot}$ maybe considered as a noncommutative de Rham complex
(cf.[10,11]). It gives a nice description of topological conformal field
theories by the LCFA, which are simultaneously such complexes. The ghost field
corresponds to "internal derivatives" in the complex.

\subhead 2.3. String backgrounds \endsubhead
Let's define string backgrounds according to [19,38]. Namely, the string
background is the set of representations of a chain complex
$C_{\Cdot}\Mor(\CA,\CB)$ in the spaces
$\Hom(H^{\otimes\#\CA},H^{\otimes\#\CB})$, where $H$ is a complex (a graded
vector space with a differential $Q$: $Q^2=0$). These representations are
morphisms of complexes and the gluings are transformed into the compositions
of operators.

As it was shown in [19,38] a string background defines a structure of a
topological QPFT--operator algebra in the space $H$.

In view of the equivalency of the topological conformal field theories and
$N\!=\!2$ superconformal ones a string background maybe transformed into the
representation of the category $\Train(\NS)$ of $N\!=\!2$ superconformal
Riemann surfaces, the train related to the Neveu--Schwarz Lie superalgebra
(it was marked in [38]).
\newpage

\head 3. Infinite dimensional noncommutative geometry of\linebreak
a self--interacting string field \endhead

\redefine\BP{\operatorname{BP}}
\define\SI{\operatorname{SI}}
\define\GI{\operatorname{GI}}
\define\GM{\operatorname{GM}}
\redefine\FG{\operatorname{FG}}
In this chapter we shall work with the following objects [1]:
\roster
\item"$\CQ$" (or the dual $\CQ^*$) --- the space of external degrees of a
freedom
of a string. The coordinates $x^\mu_n$ on $\CQ$ are the Taylor coefficients of
functions $x^\mu(z)$, which determines a world-sheet of a string in a
complexified target space.
\item"$M(\Vir)$" --- the space of internal degrees of a freedom of a string,
which is identified via Kirillov construction with the class $S$ of the
univalent functions $f(z)$; the natural coordinates on $S$ are coefficients
$c_k$ of the Taylor expansion of an univalent function $f(z)$:
$f(z)=z+c_1z^2+c_2z^3+c_3z^4+\ldots+c_nz^{n+1}+\ldots$.
\item"$\CC$" --- the universal deformation of a complex disc with $M(Vir)$
as a base and with fibers isomorphic to $D_+$; the coordinates on $\CC$ are
$z, c_1, c_2, \ldots c_n,\ldots$, where $c_k$ are coordinates on the base and
$z$ is a coordinate in the fibers.
\item"$M(\Vir)\ltimes\CQ^*$" --- the space of both external and internal
degrees of freedom of a string, the same as the bundle over $M(\Vir)$
associated with $p:\CC\mapsto M(\Vir)$, which fibers are $\Map(\Cal
C/M(\Vir);\AC^n)^*$ --- linear spaces dual to ones of mappings of fibers of
$p:\CC\mapsto M(\Vir)$ into $\AC^n$.
\item"$\Omega^{\SI}_{\BP}(E_{h,c})$" --- the space of {\it the Banks--Peskin
differential forms}, they are some geometric objects on $M(\Vir)\cdot\CQ^*$.
\item"$Q$" --- {\it the natural (geometric) BRST--operator\/} in
$\Omega^{\SI}_{\BP}(E_{h,c})$; $Q^2=0$ iff $c=26$.
\item"$\Omega^{\SI}_{\BP}(E^*_{h,c})^*$" --- the space of {\it the Siegel
string fields} with the (pseudo)hermitean metric $(\cdot|\cdot)$
\item"$Q^*$" --- {\it the Kato--Ogawa BRST--operator\/} in the space of Siegel
string
fields, the conjugate to $Q$; it defines a new (pseudo)hermitean metric
$((\cdot|\cdot))=(\cdot|\,Q^*\,|\cdot)$ in $\Omega^{\SI}_{\BP}(E^*_{h,c})$.
\item"$\FG_{h,c}(M(\Vir))$" --- {\it the Fock--plus--ghost bundle\/} over
$M(\Vir)$, its sections are just the Banks--Peskin differential forms.
\item"$\nabla^{\GM}$" --- {\it the Gauss--Manin string connection\/} in
$\FG_{h,c}(M(\Vir))$; the covariantly constant sections of which are {\it the
Bowick--Rajeev vacua}.
\item"$D_{\nabla^{\GM}}$" --- the covariant differential with respect to the
Gauss--Manin string connection.
\item"$\Omega^{\SI}_{\BP}(E_{h,c}^*)^*_{\GI}$" --- the space of {\it the
gauge--invariant Siegel string fields}, it is just the dual to the space of the
Bowick--Rajeev vacua; this space possess a (pseudo)hermitean metric
$(\cdot,\cdot)_0$, which is a restiction of the metric $(\cdot,\cdot)$.
\item"$Q^*_0$" -- the Kato--Ogawa BRST--operator in the space of
gauge--invariant Siegel string fields ($Q^*=D_{\nabla^{\GM}}+Q^*_0$);
the (pseudo)hermitean metric $((\cdot|\cdot))=(\cdot|\,Q^*_0\,|\cdot)$ is just
the
restriction of $((\cdot|\cdot))$ on $\Omega^{\SI}_{\BP}(E_{h,c}^*)^*_{\GI}$.
\endroster

It should be mentioned that the spaces of the Banks--Peskin differential
forms, the Siegel string fields, the gauge--invariant Siegel string fields,
the Bowick--Rajeev vacua are indeed superspaces and various objects on them
are (odd or even) superobjects, but the prefix 'super' will be omitted
everywhere.

The action of the Virasoro algebra $\Cvir$ in the space of the Banks--Peskin
differential forms in the flat background have the form [1]
$$\allowdisplaybreaks\align
L_p=&\fp{c_p}+\sum_{k\ge
1}(k+1)c_k\fp{c_{k+p}}-\sum_k(p+2k)\xi_{k+p}\fp{\xi_k}\quad (p>0),\\
L_0=&\sum_{k\ge 1}kc_k\fp{c_k}+\sum_{k\ge
1}kx^\mu_k\fp{x^\mu_k}-2\sum_kk\xi_k\fp{\xi_k}+h,\\
L_{-1}=&\sum_{k\ge 1}((k+2)c_{k+1}-2c_1c_k)\fp{c_k}+2c_1\sum_{k\ge
1}kx_k^\mu\fp{x^\mu_k}+\\&\sum_{k\ge
1}(k+1)x^\mu_{k+1}\fp{x^\mu_k}+\sum_k(1-2k)\xi_{k-1}\fp{\xi_k}+
2hc_1,\\
L_{-2}=&\sum_{k\ge 1}((k+3)c_{k+2}-(4c_2-c_1^2)c_k-b_k(c_1,\ldots
c_{k+2}))\fp{c_k}+\\&(4c_2-c^2_1)\sum_{k\ge 1}kx_k^\mu\fp{x^\mu_k}+
3c_1\sum_{k\ge 1} x^\mu_{k+1}\fp{x^\mu_k}+\\&\sum_{k\ge
1}(k+2)x^\mu_{k+2}\fp{x^\mu_k}+
\sum_k2(1-k)\xi_{k-2}\fp{\xi_k}+\\&\frac{x^2_1}2+h(4c_2-c_1^2)+
\frac{c}2(c_2-c_1^2),\\
L_{-n}=&\frac1{(n-2)!}\ad^{n-2}L_{-1}\cdot L_{-2}\quad (n>2),
\endalign
$$
(here $e_\mu=0$).

The action of the Virasoro algebra $\Cvir$ in the space of the Siegel string
fields in the flat background have the form [1]
$$\allowdisplaybreaks\align
L_{-p}=&\,c_p+\sum_{k\ge
1}(k+1)c_{k+p}\fp{c_k}-\sum_k(k+p)\xi^*_{k-p}\fp{\xi^*_k}\quad (p>0),\\
L_0=&\sum_{k\ge 1}kc_k\fp{c_k}+\sum_{k\ge
1}kx^\mu_k\fp{x^\mu_k}-\sum_k\xi^*_k\fp{\xi^*_k}+h,\\
L_1=&\sum_{k\ge 1}c_k((k+2)\fp{c_{k+1}}-2\fp{c_1}\fp{c_k})+2\sum_{k\ge
1}kx^\mu_k\fp{c_1}\fp{x^\mu_k}+\\
&\sum_{k\ge 1}(k+1)x^\mu_k\fp{x^\mu_{k+1}}+\sum_k(1-k)\xi^*_{k+1}\fp{\xi^*_k}+
2h\fp{c_1},\\
L_2=&\sum_{k\ge
1}c_k((k+3)\fp{c_{k+2}}-(4\fp{c_2}-(\fp{c_1})^2)\fp{c_k}-
b_k(\fp{c_1},\ldots\fp{c_{k+2}}))+\\&\sum_{k\ge
1}kx^\mu_k\fp{x^\mu_k}(\fp{c_2}-(\fp{c_1})^2)+3\sum_{k\ge
1}(k+1)x^\mu_k\fp{c_1}\fp{x^\mu_{k+1}}+\\
&\sum_{k\ge 1}(k+2)x^\mu_k\fp{x^\mu_{k+2}}+\sum_k(2-k)\xi^*_{k+2}\fp{\xi^*_k}+
\\&\frac12\frac{\partial^2}{\partial
x^2_1}+h(4\fp{c_2}-(\fp{c_1})^2)+\frac{c}2(\fp{c_2}-
(\fp{c_1})^2),\\
L_n=&\frac{(-1)^n}{(n-2)!}\ad^{n-2}L_1\cdot L_2\quad (n>2).
\endalign
$$
Here $\xi^*_k$ and $\xi_k$ are ghosts and antighosts, respectively.

The formulas for a curved background maybe received from par.2.2. of [1].

\subhead 3.1. Remarks on infinite dimensional geometry of a free string field:
Bowick--Rajeev string instantons on curved backgrounds, K\"ahler structures
and Poisson brackets on instanton spaces, background (in)depen\-dence
\endsubhead
It should be mentioned that the gauge--invariant Siegel string fields (as well
as the Bowick--Rajeev vacua) exist if and only if the Gauss--Manin connection
$\nabla^{\GM}$ is flat (more precisely, flat on a sufficiently large subbundle
--- see [1]). The conditions of a flatness of the Gauss--Manin connection pick
out the critical dimension and produce a certain equation on the background
metric field ({\it string Einshtein equations\/}).  So the formalism based on
the Bowick--Rajeev vacua is essentially set on the moduli space of solutions
of the string Einshtein equations. Indeed, it is necessary to consider a
string field theory on arbitrary metric backgrounds for the applications to
the differential geometry (of course, taking an importance of string Einshtein
equations in account). Similar situation is appeared in ordinary differential
geometry, where all constructions have a general meaning, but Einshtein
manifolds also play a considerable role.

A generalisation maybe performed in a standard "instantonic" way. Namely, one
can consider minima of a certain functional ("action") instead of
Bowick--Rajeev vacua. Namely, such string instanton action maybe chosen of the
form $S=
S\left(\frac{(D^*_{\nabla^{\GM}}\Phi,D^*_{\nabla^{\GM}}\Phi)}
{(\Phi,\Phi)}\right)$, $S\ge 0$, $S(0)=0$
(it is clamed the metric $(\cdot|\cdot)$ to be hermitean), where $\Phi$ is a
Siegel string field. A minimum of the action $S$ will be called {\it the
Bowick--Rajeev instanton}. We shall {\sl suppose} that such minima exist.

\redefine\BR{\operatorname{BR}}
Let us mentioned that the space $\BI_{\BR}=\BI_{\BR}(g_{\mu\nu};h,c)$
($g_{\mu\nu}$ is a background metric and $h, c$ are parameters of
$\Omega_{\BP}^{\SI}(E^*_{h,c})^*$) of the Bowick--Rajeev instantons is a curved
CR--space in contrast to a case of the Bowick--Rajeev vacua. Nevertheless,
one may consider a restriction of the hermitean metric $((\cdot|\cdot))$ on
$\BI_{\BR}$ supplying it by a CR-(pseudo)-K\"ahler form $w_{g_{\mu\nu};h,c}$.

\define\isotr{\operatorname{\sssize isotr}}
Some remarks are necessary. First, it should be mentioned that one needs
rather in the Poisson brackets than in the CR-(pseudo)-K\"ahler structure for a
quantization. If the CR-(pseudo)-K\"ahler metric $w_{g_{\mu\nu};h,c}$ is
nondegenerate (it claims, in particular, that the tangent space to $\BI_{\BR}$
does not contain a solution of the equation $Q^*\Phi=0$) then a transition to
Poisson brackets is straightforward but it is not so in general. Second, the
Bowick--Rajeev instantons are certain Siegel string fields, whereas in a flat
background the correct Poisson brackets are defined in functionals on the
space of the Banks--Peskin differential forms.  With respect to the 2nd remark
one should mentioned that the metric $(\cdot,\cdot)$ is nondegenerate so the
spaces of Siegel string fields and the Banks--Peskin differential forms maybe
identified. With respect to the 1st remark one may use the following rather
standard construction. Namely, let us consider the isotropic foliation $\Cal
F_{\isotr}$ on $\BI_{\BR}$ and a relative cotangent bundle, which will be
called {\it the $\Pi$--instanton space\/} (to emphasize an analogy with
$\Pi$--spaces considered in [1]) and denoted by $\Pi\BI_{\BR}$. The additional
coordinates on $\Pi\BI_{\BR}$ maybe considered as {\it the extraghosts\/}.
Then the $\Pi$--instanton space is symplectic and therefore the Poisson
brackets in functionals on it are defined.

There are two possibilities to get rid of extraghosts. First, one may exclude
them considering a flat connection in the relative cotangent bundle and
performing a hamiltonian reduction [40]. But it seems that hamiltonian
reduction maybe performed only in degenerate cases in view of nontrivial
topology of $\BI_{\BR}$. Second, one may suppose that free string field theory
on a curved background have no a quasiclassical counterpart and is analogous
to minimal models in quantum conformal field theory. In this case one is
willing to perform something analogous to the Alekseev--Shatashvili
construction [41] (see also comments in par.1.4. of [1]) and here an analogy
between $\Pi$--instanton spaces and $\Pi$--spaces will be crucial.  This is a
more reasonable but only hypothetical way and, therefore, we have to work
with extraghosts in free string field theory.

The topological properties of $\Pi\BI_{\BR}$ (the topology of the
$\Pi$--instanton space $\Pi\BI_{\BR}$ is determined, first, by a topology of
the Bowick--Rajeev instanton space $\BI_{\BR}$ itself, second, by a topology
of the isotropic foliation $\CF_{\isotr}$ on it) are essential characteristics
of a metric background. If they differ it implies the non--equivalency of free
string field theories for considered backgrounds. It should be marked that
free string field theories are determined by pair
$(\Pi\BI_{\BR};\{\cdot,\cdot\})$, where $\{\cdot,\cdot\}$ are Poisson brackets
in $\CO(\Pi\BI_{\BR})$. If such pairs are the same the theories are
equivalent. If the considered backgrounds are solutions of string Einshtein
equations then topological equivalence of vacua spaces implies the equivalency
of theories. So the theories from one connected component of the moduli space
of solutions of the string Einshtein equations are equivalent, i.e. an
infinitesimal background independence holds. Both facts (an infinitesimal
background independence and that topological equivalence implies equivalence
of theories) do not hold, in general, for arbitrary backgrounds. Indeed,
nondegenerate Poisson brackets $\{\cdot,\cdot\}$ define a CR-K\"ahler form
$\Pi\BI_{\BR}$, which is interesting as an element of $H^{2}(\Pi\BI_{\BR})$.
If cohomology $H^{2}(\Pi\BI_{\BR})$ of the $\Pi$--instanton space
$\Pi\BI_{\BR}$ is not trivial then the cohomological classes are invariants of
free string field theories. Therefore, in general, background dependence is
described by the CR--differential topology of the $\Pi$--instanton space
$\Pi\BI_{\BR}$ and the cohomological class from $H^{2}(\Pi\BI_{\BR})$.

It should be marked, however, that a special choice of the instanton action
$S$ may essentially simplify a picture. For example, if one is able to choose
$S(t)=t$ then $\BI_{\BR}$ is a linear subspace of
$\Omega^{\SI}_{\BP}(E_{h,c})$ (or $\Omega^{\SI}_{\BP}(E^*_{h,c})^*$),
therefore $\Pi\BI_{\BR}$ as well as $\BI_{\BR}$ are contractible, moreover
$\Pi\BI_{\BR}$ is a flat CR--space, which carries a constant Poisson brackets.
Hence, two theories are equivalent iff the corresponding pairs $(\BI_{\BR},
\CF^{\isotr})$ are isomorphic as pairs of topological vector spaces. If one
prefers to deal with them as Hilbert spaces then {\it the full metric
background independence\/} of free string field theory will be received.

\subhead 3.2. String field algebras, Poisson brackets on duals to them
and their quantization (flat background)\endsubhead
Above we have deal with a chiral case but we need in a real one below. They
differs by minor "second order" details, so we will not repeat all
constructions again. In this paragraph we shall consider a flat background.

\define\sf{\operatorname{\bold s\bold f}}
\define\enl{\operatorname{enl}}
First of all, it should be mentioned that the space of the Siegel string
fields do not admit a structure of local conformal field algebra because it
contains Verma modules over the Virasoro algebra $\Cvir$ only of a discrete
spectrum of weights. To get rid of this difficulty let's enlarge the spaces of
the Banks--Peskin differential forms and the Siegel string fields. The space
of enlarged Banks--Peskin differential forms will be defined as the space
$\Omega^{\cdot}_{\BP;\enl}=\Omega^{\cdot}(\tilde{\Bbb
C}^*,\Omega^{\SI}_{\BP}(E_{h,c}))$. Let $t$ be a coordinate on $\tilde{\Bbb
C}^*$ then the action of the Virasoro algebra generators in
$\Omega^i_{\BP;\enl}$ ($i=1,2$) will be defined by the formulas
$$\allowdisplaybreaks\align
L_p=&\fp{c_p}+\sum_{k\ge
1}(k+1)c_k\fp{c_{k+p}}-\sum_k(p+2k)\xi_{k+p}\fp{\xi_k}\quad (p>0),\\
L_0=&\sum_{k\ge 1}kc_k\fp{c_k}+\sum_{k\ge
1}kx^\mu_k\fp{x^\mu_k}-2\sum_kk\xi_k\fp{\xi_k}-t\fp{t}-i+h,\\
L_{-1}=&\sum_{k\ge 1}((k+2)c_{k+1}-2c_1c_k)\fp{c_k}+2c_1\sum_{k\ge
1}kx_k^\mu\fp{x^\mu_k}+\\&\sum_{k\ge
1}(k+1)x^\mu_{k+1}\fp{x^\mu_k}+\sum_k(1-2k)\xi_{k-1}\fp{\xi_k}+t^2\fp{t}+2it+
2hc_1,\\
L_{-2}=&\sum_{k\ge 1}((k+3)c_{k+2}-(4c_2-c_1^2)c_k-b_k(c_1,\ldots
c_{k+2}))\fp{c_k}+\\&(4c_2-c^2_1)\sum_{k\ge 1}kx_k^\mu\fp{x^\mu_k}+
3c_1\sum_{k\ge 1} x^\mu_{k+1}\fp{x^\mu_k}+\\&\sum_{k\ge
1}(k+2)x^\mu_{k+2}\fp{x^\mu_k}+
\sum_k2(1-k)\xi_{k-2}\fp{\xi_k}-3t^3\fp{t}+\\&\frac{x^2_1}2+3it^2+h(4c_2-c_1^2)+
\frac{c}2(c_2-c_1^2),\\
L_{-n}=&\frac1{(n-2)!}\ad^{n-2}L_{-1}\cdot L_{-2}\quad (n>2).
\endalign
$$
The space of enlarged Siegel string fields will be defined in analogous way
$\Omega^{\cdot}_{\sf;\enl}=\Omega^{\cdot}(\tilde{\Bbb
C}^*,\Omega^{\SI}_{\BP}(E_{h,c}^*)^*)$. The action of the Virasoro algebra
generators in $\Omega^i_{\sf;\enl}$ ($i=1,2$) will be defined by the formulas
$$\allowdisplaybreaks\align
L_{-p}=&\,c_p+\sum_{k\ge
1}(k+1)c_{k+p}\fp{c_k}-\sum_k(k+p)\xi^*_{k-p}\fp{\xi^*_k}\quad (p>0),\\
L_0=&\sum_{k\ge 1}kc_k\fp{c_k}+\sum_{k\ge
1}kx^\mu_k\fp{x^\mu_k}-\sum_k\xi^*_k\fp{\xi^*_k}+t\fp{t}+i+h,\\
L_1=&\sum_{k\ge 1}c_k((k+2)\fp{c_{k+1}}-2\fp{c_1}\fp{c_k})+2\sum_{k\ge
1}kx^\mu_k\fp{c_1}\fp{x^\mu_k}+\\
&\sum_{k\ge 1}(k+1)x^\mu_k\fp{x^\mu_{k+1}}+\sum_k(1-k)\xi^*_{k+1}\fp{\xi^*_k}+
+t^2\fp{t}+2it+2h\fp{c_1},\\
L_2=&\sum_{k\ge
1}c_k((k+3)\fp{c_{k+2}}-(4\fp{c_2}-(\fp{c_1})^2)\fp{c_k}-
b_k(\fp{c_1},\ldots\fp{c_{k+2}}))+\\&\sum_{k\ge
1}kx^\mu_k\fp{x^\mu_k}(\fp{c_2}-(\fp{c_1})^2)+3\sum_{k\ge
1}(k+1)x^\mu_k\fp{c_1}\fp{x^\mu_{k+1}}+\\
&\sum_{k\ge
1}(k+2)x^\mu_k\fp{x^\mu_{k+2}}+\sum_k(2-k)\xi^*_{k+2}\fp{\xi^*_k}+t^3\fp{t}+
\\&\frac12\frac{\partial^2}{\partial
x^2_1}+3it^2+h(4\fp{c_2}-(\fp{c_1})^2)+\frac{c}2(\fp{c_2}-
(\fp{c_1})^2),\\
L_n=&\frac{(-1)^n}{(n-2)!}\ad^{n-2}L_1\cdot L_2\quad (n>2).
\endalign
$$
One may also construct the enlarged BRST--operators $Q_{\enl}$ and
$Q_{\enl}^*$ as exterior differentials in $\Omega^{\cdot}_{\BP;\enl}$ and
$\Omega^{\cdot}_{\sf;\enl}$ from old BRST--operators $Q$ and $Q^*$.

\proclaim{Proposition 7} The space $\Omega^{\cdot}_{\sf;\enl}$ admits a
structure of a BRST--invariant LCFA, covariant with respect to the
Gauss--Manin connection $\nabla_{\GM}$.
\endproclaim

\proclaim{Lemma} The space $\Omega^0_{\sf;\enl}$ admits a structure of a LCFA,
covariant with respect to the Gauss--Manin connection.
\endproclaim

\demo{Proof of lemma} One should apply the procedure of the renormalization of
a
pointwise product of par.1.3. to the standard vertex operator algebra (of the
first quantized string on a flat background) in the
space $\Omega^{\SI}_{\BP}(E^*_{h,c})^*_{\GI}$.
\enddemo

To obtain a structure of a BRST--invariant LCFA, covariant with respect to the
Gauss--Manin connection $\nabla_{\GM}$ in the whole space
$\Omega^{\cdot}_{\sf;\enl}$ one should use a construction of the Prop.6.

Therefore, the space $\Omega^{\cdot}_{\sf;\enl}$ maybe considered as a
noncommutative de Rham complex with respect to the enlarged BRST--operators.
Such complex will be called {\it the enlarged string field algebra}.

Now one may apply the construction of par.1.3. to receive a structure of {\it
the non--associative string field algebra\/} in the space of Siegel string
fields (cf. [42]). The point projectors $P_t$ map $\Omega^0_{\sf;\enl}$ onto
$\Omega^{\SI}_{\BP}(E^*_{h,c})^*$ and the point imbeddings $I_t$ realize the
inversed mappings from $\Omega^{\SI}_{\BP}(E^*_{h,c})^*$ to
$\Omega^0_{\sf;\enl}$.

\define\wit{\operatorname{^{\circlearrowleft}\Fw\Fi\Ft}} It should be
mentioned that elements of $\Omega^0_{\sf;\enl}$ form a Lie algebra under a
commutator. This Lie algebra admits a central extension by the imaginary part
of $((\cdot|\cdot))$. If one consider the space of connections on $\tilde{\Bbb
C}^*$ valued in $\Omega^{\SI}_{\BP}(E_{h,c}^*)^*$, i.e. gauge fields on
$\tilde{\AC}^*$ valued in Siegel string fields, then elements of
$\Omega^0_{\sf;\enl}$ will realize their infinitesimal gauge--transformations.
These gauge--transformations are closed (as in Witten string field theory
[28]) so the corresponding Lie algebra will be called Witten string Lie
algebra and will be denoted by $\wit$ (the circled arrow $\circlearrowleft$ is
a code for "string"). The space of $\nabla^{\GM}$--covariant elements of
$\wit$ will be denoted by $\wit_{\nabla^{\GM}}$ and also called by the Witten
string Lie algebra.

There are defined canonical (Lie--Berezin) Poisson brackets in the
space $\wit^*$ (or $\wit^*_{\nabla^{\GM}}$) dual to the Witten string algebra
$\wit$ (or $\wit_{\nabla^{\GM}}$, which may be quantized as such.

\define\zwie{\operatorname{^{\circlearrowleft}\Fz\Fw\Fi\Fe}} Let's mention
that the point projectors and the point imbeddings allow to perform a
hamiltonian reduction of these Poisson brackets and to receive the
nonpolynomial Poisson brackets in the space of functionals on the
Banks--Peskin differential forms (or Bowick--Rajeev vacua). These Poisson
brackets generate a Lie quasi(pseudo)algebra (quasialgebra in terminology of
[43] and pseudoalgebra in terminology of [40]) of non--polynomial
infinitesimal gauge--transformations. The non--polynomial transformations in
the space of Bowick--Rajeev vacua were considered in [44;38]. The Lie
quasi(pseudo)algebra generated by them will be called the Zwiebach string Lie
quasi(pseudo)algebra and will be denoted by $\zwie_{\nabla^{\GM}}$, whereas
the corresponding Lie quasi(pseudo)algebra in the space of Banks--Peskin
differential forms will be denoted by $\zwie$. The nonpolynomial Poisson
brackets are defined in functionals on the dual $\zwie^*$ (or
$\zwie^*_{\nabla^{\GM}}$) to the Lie quasi(pseudo)al\-ge\-b\-ra $\zwie$ (or
$\zwie_{\nabla^{\GM}}$). It should be mentioned that the central charge of the
Witten string Lie algebra is transformed by hamiltonian reduction into the
inverse $\gg^{-1}$ of the coupling constant $\gg$; this fact enlight a
non--perturbative nature of a coupling constant $\gg$. Also it should be
marked that the Zwiebach string Lie quasi(pseudo)algebra maybe received from
the non--associative string field algebra as its "commutator" algebra. More
precisely, the higher operations of Sabinin--Mikheev multialgebra [45]
constructed by the Zwiebach string Lie algebra are just higher commutators in
the non--associative string field algebra.

Thus, the nonpolynomial string field theory maybe received from Witten-type
string field theory in enlarged space by a hamiltonian reduction.

Some remarks on quantization are necessary. There are two possibilities to
quantize the nonpolynomial Poisson brackets on $\zwie^*_{\nabla^{\GM}}$.
First, one may quantized them as nonlinear brackets asymptotically [40].
Second, one may perform a quantum reduction of quantized brackets on
$\wit^*_{\nabla^{\GM}}$ (i.e. to receive the corresponding algebra of
observables as a certain quantum reduction of $\CU(\wit_{\nabla^{\GM}})$).

It is necessary to mark that the objects, which we were constructed describe
the interacting string field theory on a tree level (or "classical interacting
string field theory" [37]). To describe this theory completely in the
nonperturbative mode one may use the following result.

\proclaim{Proposition 8} The Witten string Lie algebra $\wit_{\nabla^{\GM}}$
(or
$\wit$) admits a structure of a Lie bialgebra.
\endproclaim

Indeed, the Siegel string field possess a structure of a string background
[38] so the enlarged string field algebra is a crossing--algebra (it holds also
for the subalgebra of $\nabla^{\GM}$--covariant elements in it).

Therefore, on the quantum level one have to consider a quantum universal
envelopping algebra $\CU_q(\wit_{\nabla^{\GM}})$ (or $\CU_q(\wit)$)
(cf.[46]) and its reductions. {\sl To construct explicitely all these objects
is a problem.}

\subhead 3.3. Remarks on classical interacting string field theory
in curved backgrounds; background (in)dependence \endsubhead
The main idea how to formulate the interacting string field theory in curved
background is to use the fact that the complex $\Omega^{\cdot}_{\sf;\enl}$
does not depend on a background, so one may consider the enlarged string field
algebra as universal for all backgrounds. That means that string interactions
(vertices and covertices) after the accounting of the internal degrees of
freedom of a string do not depend on metric background. Therefore, the Witten
string Lie algebra $\wit$ as well as the Zwiebach string Lie
quasi(pseudo)algebra $\zwie$ are also universal so all information on the
metric background is encoded in the Gauss--Manin connection $\nabla^{\GM}$.
For a general background this connection is not flat in any sense so the
algebras $\wit_{\nabla^{\GM}}$ and $\zwie_{\nabla^{\GM}}$ do not exist. And if
they exist (f.e. for solutions of string Einshtein equations) they are not, in
general, identical.

Let's combine the approaches of par.3.1. and par.3.2. for an arbitrary metric
background.

Let' consider an arbitrary (pseudo)--K\"ahler symplectic leaf $\CO_{\Bp}$
($\Bp\in\bold P$ pa\-ra\-me\-t\-ri\-zes such leaves) of the non--polynomial
brackets on
$\Omega^{\SI}_{\BP}(E_{h,c})$ (so $\CO_{\Bp}$ is na\"\i vely a "coadjoint
orbit" of the Zwiebach string Lie algebra $\zwie$ and $\bold P$ is the space
of such orbits). Also let us consider the intersection
$\BI_{\BR}(\Bp)=\BI_{\BR}\cap\CO_{\Bp}$ of this leaf with the Bowick--Rajeev
instanton space $\BI_{\BR}$; the (pseudo)--K\"ahler structure on $\CO_{\Bp}$
defines a CR--(pseudo)--K\"ahler metric on $\BI_{\BR}(\Bp)$. One may now use a
construction of par.3.1. and build the $\Pi$--instanton space
$\Pi\BI_{\BR}(\Bp)$ and Poisson brackets in functionals on it.
So the classical interacting string field theory is determined by the
following data:
\roster
\item"--" the $\Pi$--instanton space $\Pi\BI_{\Bp}$;
\item"--" the Poisson brackets (or (pseudo)--K\"ahler metric) on it.
\endroster
The $\Pi$--instanton space $\Pi\BI_{\Bp}$ is constructed from
\roster
\item"--" the Bowick--Rajeev instanton space $\BI_{\BR}(\Bp)$;
\item"--" the isotropic foliation $\CF_{\isotr}$ on it.
\endroster
The Bowick--Rajeev instanton space $\BI_{\BR}(\Bp)$ is an intersection of
\roster
\item"--" the fixed (pseudo)-K\"ahler symplectic leaf $\CO_{\Bp}$ of the
background independent Poisson brackets on $\Omega^{\SI}_{\BP}(E_{h,c})$;
\item"--" the Bowick--Rajeev instanton space $\BI_{\BR}$, which does not
depend on the coupling constant $\gg$.
\endroster
The isotropic foliation $\CF_{\isotr}$ depends as on background as on $\gg$.

So the independent parameters of classical interacting string field theory are
(1) the background metric $g_{\mu\nu}$, (2) the parameters of string fields
$h$ and $c=26$, (3) the coupling constant $\gg$, (4) the value of coordinate
$\Bp$ on the space $\bold P$ of (pseudo)--K\"ahler symplectic leaves of the
non--polynomial Poisson brackets on the space of Banks--Peskin differential
forms.

It was proved in [46] that the classical interacting string field theory is
infinitesimally independent on the metric background on the space of solutions
of string Einshtein equations. That means that in this case infinitesimal
variations of the Gauss--Manin connection does not influence on the resulting
string field theory.

\define\integral{\operatorname{\sssize integral}\tsize}
\subhead 3.4. Open problems of the nonperturbative approach on a quantum
level \endsubhead
There are three main open problem to construct the nonperturbative quantum
interacting string field theory:
\roster
\item"(1)" to quantize the space of parameters; i.e. to extract for a fixed
$h$ and $\gg$ the set $\bold P^{\integral}=\{\Bp\!\in\!\bold P:$ the
CR--(pseudo)--K\"ahler metric on $\Pi\BI_{\BR}(\Bp)$ belongs to
$H^{2}(\Pi\BI_{\BR}(\Bp);\AZ)$;
\item"(2)" to find the selection rules, which extract triples $(h,\gg,\Bp)$
for which the theories are unitarizable, i.e. a hermitean metric in a Fock
space over $\Pi\BI_{\BR}(\Bp)$ is non--negatively definite;
\item"(3)" to find the quantum Clebsch--Gordan coefficients, which define
tensor products of these Fock spaces for different $\Bp$ and maybe $h$ ($\gg$
are fixed).
\endroster

Also an important problem is a quantum metric background (in)dependence in the
case of arbitrary curved backgrounds. Such infinitesimal independence for the
metrics obeying string Eishtein equations was proved in [48] in context of the
Lagrangian formulation of the perturbative quantum interacting string field
theory.

Another problem is to compare the results of perturbative and nonperturbative
approaches and to estimate the nonperturbative effects.

\head Conclusions \endhead

So an infinite dimensional geometric interpretation of a self--interacting
string field theory was given and relations between various approaches to the
second quantization of an interacting string were described in terms of the
geometric quantization. A certain algorythm to construct a quantum
nonperturbative interacting string field theory in the quantum group
formalism was proposed; the main problems were formulated in geometric terms.

Their more mathematically rigorous and detailed discussion will be contained
in the forthcoming papers. Applications of the developped machinery to
problems of differential geometry will be also discussed there.

\Refs
\roster
\item"[1]" Juriev D., Infinite dimensional geometry and quantum
field theory of strings. I. Infinite dimensional geometry of a second
quantized free string. Alg. Groups Geom. 11(2) (1994) 127-180; hep-th/9403068.
\item"[2]" Juriev D., Infinite dimensional geometry and quantum field theory
of strings. III. Infinite dimensional $W$--geometry of a second quantized free
string. Preprint LPTENS 93/53; hep-th/9401026.
\item"[3]" Morozov A.Yu., Strings -- what are they? Soviet Phys. Uspekhi 35
(1992) 671--714; Integrability and matrix models. Preprint ITEP--M2/93 and ITFA
93--10; hep-th/9303139.
\item"[4]" Juriev D., Quantum conformal field theory as infinite dimensional
noncommutative geometry, Russian Math. Surveys 46(4) (1991) 135--163. (see
also Juriev D., Theor. Math. Phys. 93(1) (1992) 1101-1105).
\item"[5]" Witten E., Two--dimensional gravity and intersection theory on
moduli spaces. Surveys in Diff. Geom. 1 (1991) 243-310; Kontsevich M.L., Manin
Yu.I., Gromov--Witten classes, quantum cohomology and enumerative geometry;
hep-th/9402147.
\item"[6]" Vafa C., Topological mirrors and quantum rings. In {\it Essays in
mirror symmetry}. Ed. S.-T.Yau, 1992; Witten E., Mirror manifolds and
topological field theory. In {\it Essays in mirror symmetry}. Ed. S.-T.Yau,
1992; Sadov V., On equivalency of Floer's and quantum cohomology. Preprint
HUTP--93/A027; hep-th/93110153; Piunikhin S., Quantum and Floer cohomology
have the same ring structure. Preprint MIT--MATH--9401.
\item"[7]" Juriev D., The vocabulary of geometry and harmonic analysis on the
infinite-dimensional manifold $\Diff_+(\circle)/\circle$. Adv. Soviet Math. 2
(1991) 233-247; A model of Verma modules over the Virasoro algebra. St.
Petersburg Math. J. 2 (1991) 401-417; Infinite dimensional geometry of the
universal deformation of a complex disk. Russian J. Math. Phys. 2 (1994);
funct-an/9401003.
\item"[8]" Juriev D., On the univalency of regular functions. Annali di
Matem. Pura ed Appl. (4) 164 (1993) 37-50.
\item"[9]" Morozov A.Yu., Perelomov A.M., Complex geometry and string theory.
Current Probl. Math., Modern Achievements, VINITI, Moscow [in Russian];
Morozov A.Yu., Perturbative methods in string theory. Elem. Part. Atom. Nucl.
23:1 (1992), 174-238 [in Russian].
\item"[10]" Connes A., {\it Introduction \`a la g\'eom\'etrie non commutative}
(InterEditions, Paris, 1990).
\item"[11]" Manin Yu.I., {\it Topics in non-commutative geometry}. Princeton
Univ. Press, Princeton, NJ, 1991.
\item"[12]" Neretin Yu.A., A complex semigroup that contains the group of
diffeomorphisms of the circle. Funct. Anal. Appl. 21 (1987) 160-161;
Holomorphic extensions of the representations of the group of diffeomorphisms
of a circle. Math. USSR--Sbornik 67 (1990) 75-97 (see also Neretin Yu.A.,
Funct. Anal. Appl. 23 (1989) 196-206).
\item"[13]" Neretin Yu.A., Infinite dimensional groups, their mantles, trains
and representations. Adv. Soviet Math. 2 (1991) 103-172 (see also Neretin
Yu.A., Thesis Doct. Sci. Diss., Steklov Math. Inst., 1992 [in Russian]).
\item"[14]" Juriev D., Quantum projective field theory: quantum--field analogs
of the Euler--Arnold equations in projective $G$--hypermultiplets. Theor. Math.
Phys. (1994).
\item"[15]" Segal G., Two-dimensional conformal field theory and modular
functors, Proc. IXth Intern. Congr. Math. Phys. (Bristol, Philadelphia). Eds.
B.Simon, A.Truman and I.M.Davies, IOP Publ. Ltd, 1989, 22-37; Gawedzki K.,
Conformal field theory. Asterisque 177--178 (1989) 95-126, Seminaire Bourbaki
1988-89, ex. 704; Moore G., Seiberg N., Classical and quantum conformal field
theory. Commun.  Math. Phys. 123 (1989) 177-254.
\item"[16]" Kirillov A.A., Juriev D.V., The K\"ahler geometry on the infinite
dimensional homogeneous manifold $M=\Diff_+(\circle)/\Rot(\circle)$. Funct.
Anal. Appl. 20 (1986) 322-324; The K\"ahler geometry on the infinite
dimensional homogeneous space $M=\Diff_+(\circle)/\Rot(\circle)$. Funct.
Anal.  Appl. 21(4) (1987) 284-293; Representations of the Virasoro algebra by
the orbit method. J. Geom. Phys. 5 (1988) 351-364 [reprinted in {\it Geometry
and Physics. Essays in honour of I.M.Gelfand}. Eds. S.G.Gindikin and
S.G.Singer. Pitagora Editrice, Bologna and Elsevier Sci. Publ., Amsterdam,
1991].
\item"[17]" Kirillov A.A., A K\"ahler structure on K-orbits of the group of
diffeomorphisms of a circle $M=\Diff_+(\circle)/\Rot(\circle)$. Funct.
Anal. Appl. 21 (1987) 122-125 (See also the elder references: Kirillov A.A.,
Funct. Anal. Appl. 15 (1981) 135-137;  Segal G., Commun. Math. Phys. 80
(1981) 301-342;  Kirillov A.A., Lect. Notes Math. 970 (1982) 101-123; In {\it
Geometry and topology in global nonlinear problems}. Voronezh, 1984, 49-67 [in
Russian]; Witten E., Commun. Math. Phys. 114 (1988) 1-53; as well as more
recent ones: Kirillov A.A., In {\it Infinite-dimensional Lie algebras and
quantum field theory}. World Scientific, Teaneck, NJ, 1988, 73-77;  In {\it
Operator algebras, unitary representations, enveloping algebras and invariant
theory}. Birkh\"auser, Boston, 1992, 73-83;  Contemp. Math. 145, AMS,
Providence, RI, 1993, 33-63).
\item"[18]" Huang Y.-Z., Geometric interpretation of vertex operator algebras.
Proc. Nat'l Acad. Sci. USA 88 (1991) 9964-9968; Applications of the geometric
interpretation of vertex operator algebras. In {\it Proc. 20th Intern. Conf.
on Diff. Geom. Methods in Theor. Phys. New York, 1991}. Eds. S.Catto and
A.Rocha. World Scientific, Singapore, 1992, V.1., 333-343;  Vertex operator
algebras and conformal field theory. Intern. J. Mod. Phys. A7 (1992)
2109-2151; Operads and the geometric interpretation of vertex operator
algebras. I. Preprint Dpt. Math. Univ. Pennsylvania 1993.
\item"[19]" Voronov A.A., Topological field theories, string backgrounds and
homotopy algebras; hep-th/9401023.
\item"[20]" Juriev D., On the representation of operator algebras of the
quantum conformal field theories. J. Math. Phys. 33 (1992) 492-496;  Algebra
$\Vrt(\Cvir;c)$ of vertex operators for the Virasoro algebra. St. Petersburg
Math. J. 3 (1992) 679-686.
\item"[21]" Bychkov S.A., Juriev D.V., Three algebraic structures of quantum
projective field theory. Theor. Math. Phys. 97(3) (1993).
\item"[22]" Juriev D., Algebraic structures of quantum projective field theory
related to fusion and braiding. Hidden additive weight. J. Math. Phys. (to
appear); hep-th/9403051.
\item"[23]" Frenkel I.B., Huang Y.-Z., Lepowsky J., On axiomatic approach to
vertex operator algebras and modules. Memoirs AMS, 104 (1993) no.594.
\item"[24]" Frenkel I., Lepowsky J., Meurman A., {\it Vertex operator algebras
and the Monster}. Acad. Press, New York, 1988; Goddard P., Meromorphic
conformal field theory. In {\it Infinite dimensional Lie algebras and groups}.
 Ed. V.G.Kac, World Scientific, Singapore, 1989, 556-587.
\item"[25]" Huang Y.-Z., Lepowsky J., Vertex operator algebras and operads.
{\it The Gelfand Mathematical Seminars, 1990-1992}.  Eds. L.Corwin, I.Gelfand
and J.Lepowsky. Birkh\"auser, Boston, 1993, 145-161;  hep-th/9301009; Operadic
formulation of the notion of vertex operator algebra. Preprint.
\item"[26]" Huang Y.-Z., Lepowsky J., Toward a theory of tensor products for
representations of a vertex operator algbera. {\it Proc. 20th Intern. Conf.
Diff. Geom. Methods in Theor. Phys}. Eds. S.Satto and A.Rocha, World
Scientific, Singapore, 1992, V.1, 344-354; A theory of tensor products for
module categories for a vertex operator algebra. I, II; hep-th/9309076,
hep-th/9309159; Tensor products of modules for a vertex operator algebra and
vertex tensor categories; hep-th/9401119.
\item"[27]" Juriev D., Local conformal field algebras. Commun. Math. Phys. 138
(1991) 569-581; 146 (1992) 427; On the structure of L-algebra $L(\Cvir)$.  J.
Funct. Anal. 101 (1991) 1-9.
\item"[28]" Witten E., Non-commutative geometry and string field theory. Nucl.
Phys. B268 (1986) 253-294.
\item"[29]" Witten E., Quantum field theory, grassmannians and algebraic
curves. Commun. Math. Phys. 113 (1988), 529-600.
\item"[30]" Krichever I.M., Novikov S.P., Algebras of Virasoro type, Riemann
surfaces and structures of the theory of solitons. Funct. Anal. Appl. 21
(1987) 126-141; Virasoro-type algebras, Riemann surfaces and strings in
Minkowsky space. Funct. Anal. Appl. 21 (1987) 294-307; Algebras of Virasoro
type, energy-momentum tensor and decomposition operators on Riemann surfaces.
Funct. Anal. Appl. 23 (1989) 19-32; Virasoro-Gelfand-Fuks type algebras,
Riemann surfaces, operator's theory of closed strings. J. Geom. Phys. 5 (1988)
631-661 [reprinted in {\it Geometry and physics. Essays in honour of
I.M.Gelfand}. Eds. S.G.Gindikin and S.G.Singer. Pitagora Editrice, Bologna and
Elsevier Sci. Publ., Amsterdam, 1991].
\item"[31]" Juriev D., Complex projective geometry and quantum projective
field theory. Theor. Math. Phys. (submitted).
\item"[32]" Huang Y.-Z., Operadic formulation of topological vertex algebras
and Gerstenhaber or Batalin-Vilkovisky algebras; hep-th/9306021.
\item"[33]" Penkava M., Schwarz A., On some algebraic structures arizing in
string theory; hep-th/9212072.
\item"[34]" Juriev D., Setting hidden symmetries free by the noncommutative
Veronese mapping; hep-th/9402130.
\item"[35]" Lian B.H., Zuckerman G.J., New perspectives on the BRST-algebraic
structure of string theory; hep-th/9211072; Getzler E., Batalin-Vilkovisky
algebras and two-dimensional topological field theory. Commun. Math. Phys. 159
(1994) 265-286.
\item"[36]" Juriev D., The explicit form of the vertex operators in
two--dimensional $\sLtwo$--invariant field theory. Lett. Math. Phys. 22 (1991)
141-144; Classification of vertex operators in two--dimensional
$\sLtwo$--invariant quantum field theory. Teor. Matem. Fiz. 86(3) (1991)
338-343 [in Russian].
\item"[37]" Zwiebach B., Closed string field theory: quantum action and the
Batalin-Vilkovisky master equations. Nucl. Phys. B390 (1993) 33-152.
\item"[38]" Kimura T., Stasheff J., Voronov A.A., On operad structures of
moduli spaces and string theory; hep-th/9307114.
\item"[39]" Lada T., Stasheff J., Introduction to sh-Lie algebras for
physicists. Preprint UNC-MATH-92/2; hep-th/9209099;  Stasheff J., Closed
string field theory, strong homotopy Lie algebras and the operad actions of
moduli spaces. Preprint UNC-MATH-93-1; hep-th/9304061.
\item"[40]" Karasev M.V. and Maslov V.P., {\it Nonlinear Poisson brackets.
Geometry and quantization}. Nauka, Moscow, 1991.
\item"[41]" Alekseev A.Yu., Shatashvili S.L., Path quantization of the
coadjoint orbits of the Virasoro group and 2d gravity. Nucl. Phys. B323
(1989), 719-733; From geometric quantization to conformal field theory.
Commun. Math. Phys. 128 (1990), 197-212.
\item"[42]" Aref'eva I.Ya., Volovich I.V., Gauge--invariant string interaction
and nonassociative algebra. Phys. Lett. B182 (1986) 159-163; String field
algebra. Phys. Lett. B182 (1986) 312-316; B189 (1987) 488.
\item"[43]" Batalin I., Quasigroup construction and first class constraints, J.
Math. Phys. 22 (1981) 1837-1850.
\item"[44]"  Saadi M., Zwiebach B., Closed string field theory from polyhedra.
Ann. Phys. (NY) 192 (1989) 213-227; Kugo T., Kunimoto H., Suehiro K.,
Non-polynomial closed string field theory. Phys. Lett. B226 (1989) 48-54;
Sonoda H., Zwiebach B., Closed string field theory loops with symmetric
factorizable quadratic differentials. Nucl. Phys. B331 (1990) 592-628; Kugo T.,
Suehiro K., Nonpolynomial closed string field theory: action and gauge
invariance. Nucl. Phys. B337 (1990) 434-466; Zwiebach B., How covariant closed
string theory solves a minimal area problem. Commun. Math. Phys. 136 (1991)
83-118; Wolf M., Zwiebach B., The plumbing of minimal area surfaces. Preprint
IASSNS-92/11; hep-th/9202062 (and references wherein).
\item"[45]" Mikheev P.O., Sabinin L.V., On the infinitesimal theory of local
loops. Soviet math. 36 (1988) 545-548.
\item"[46]" Drinfeld V.G., Quantum groups. Proc. Intern. Congr. Math.,
Berkeley, California, 1986; Reshetikhin N.Yu., Takhtadzhan L.A., Faddeev L.D.,
Quantization of Lie algebras and Lie groups. St. Petersburg Math. J. 1 (1990)
193-225.
\item"[47]" Sen A., Zwiebach B., A proof of local background independence of
classical closed string field theory. Preprint MIT-CTP-2222; hep-th/9307088.
\item"[48]" Sen A., Zwiebach B., Quantum background independence of closed
string field theory. Preprint MIT-CTP-2244; hep-th/9311009.
\endroster
\endRefs
\enddocument